# Thirteen New M Dwarf + T Dwarf Pairs Identified with WISE/NEOWISE

Federico Marocco,[1] J. Davy Kirkpatrick,[1,2] Adam C. Schneider,[3,2] Aaron M. Meisner,[4,2] Mark Popinchalk,[5,6,7] Christopher R. Gelino,[1] Jacqueline K. Faherty,[5,2] Adam J. Burgasser,[8] Dan Caselden,[5] Jonathan Gagné,[9,10] Christian Aganze,[8] Daniella C. Bardalez Gagliuffi,[11] Sarah L. Casewell,[12] Chih-Chun Hsu,[8] Rocio Kiman,[13] Peter R. M. Eisenhardt,[14] Marc J. Kuchner,[15,2] Daniel Stern,[14] Léopold Gramaize,[2] Arttu Sainio,[2] Thomas P. Bickle,[16,2] Austin Rothermich,[5,6,7] William Pendrill,[2] Melina Thévenot,[2] Martin Kabatnik,[2] Giovanni Colombo,[2] Hiro Higashimura (東村滉),[17] Frank Kiwy,[2] Elijah J. Marchese,[18] Nikolaj Stevnbak Andersen,[2] Christopher Tanner,[2] Jim Walla,[2] Zbigniew Wedracki,[2] and The Backyard Worlds Collaboration

[1] IPAC, Mail Code 100-22, Caltech, 1200 E. California Blvd., Pasadena, CA 91125, USA
[2] Backyard Worlds: Planet 9
[3] United States Naval Observatory, Flagstaff Station, 10391 West Naval Observatory Rd., Flagstaff, AZ 86005, USA
[4] NSF's National Optical-Infrared Astronomy Research Laboratory, 950 N. Cherry Avenue, Tucson, AZ 85719, USA
[5] Department of Astrophysics, American Museum of Natural History, Central Park West at 79th Street, New York, NY 10024, USA
[6] Department of Physics, Graduate Center, City University of New York, 365 5th Ave., New York, NY 10016, USA
[7] Department of Physics and Astronomy, Hunter College, City University of New York, 695 Park Avenue, New York, NY, 10065, USA
[8] Department of Physics, University of California, San Diego, 9500 Gilman Dr, La Jolla, CA, 92093, USA
[9] Planétarium Rio Tinto Alcan, Espace pour la Vie, 4801 av. Pierre-de Coubertin, Montréal, Québec, Canada
[10] Institute for Research on Exoplanets, Université de Montréal, Département de Physique, C.P. 6128 Succ. Centre-ville, Montréal, QC H3C 3J7, Canada
[11] Department of Physics & Astronomy, Amherst College, 25 East Drive, Amherst, MA 01003, USA
[12] School of Physics and Astronomy, University of Leicester, University Road, Leicester, LE1 7RH, UK
[13] Department of Astronomy, California Institute of Technology, Pasadena, CA 91125, USA
[14] Jet Propulsion Laboratory, California Institute of Technology, 4800 Oak Grove Dr., Pasadena, CA 91109, USA
[15] Exoplanets and Stellar Astrophysics Laboratory, NASA Goddard Space Flight Center, 8800 Greenbelt Road, Greenbelt, MD 20771, USA
[16] School of Physical Sciences, The Open University, Milton Keynes, MK7 6AA, UK
[17] Earl of March Intermediate School, 4 The Pkwy, Kanata, ON K2K 1Y4, Canada
[18] Independent researcher, USA

## ABSTRACT

We present the discovery of 13 new widely separated T dwarf companions to M dwarf primaries, identified using WISE/NEOWISE data by the CatWISE and Backyard Worlds: Planet 9 projects. This sample represents a ∼60% increase in the number of known M+T systems, and allows us to probe the most extreme products of binary/planetary system formation, a discovery space made available by the CatWISE2020 catalog and the Backyard Worlds: Planet 9 effort. Highlights among the sample are WISEP J075108.79-763449.6, a previously known T9 thought to be old due to its SED, which we now find is part of a common-proper-motion pair with L 34-26 A, a well studied young M3 V star within 10 pc of the Sun; CWISE J054129.32–745021.5 B and 2MASS J05581644–4501559 B, two T8 dwarfs possibly associated with the very fast-rotating M4 V stars CWISE J054129.32–745021.5 A and 2MASS J05581644–4501559 A; and UCAC3 52-1038 B, which is among the widest late T companions to main sequence stars, with a projected separation of ∼7100 au. The new benchmarks presented here are prime *JWST* targets, and can help us place strong constraints on formation and evolution theory of substellar objects as well as on atmospheric models for these cold exoplanet analogs.

Corresponding author: Federico Marocco
federico@ipac.caltech.edu

## 1. INTRODUCTION

The study of binaries and multiple stellar systems is the principal means to empirically validate models of star formation and evolution, since the distribution of physical properties (age, mass, and metallicity) and orbital parameters (separation, eccentricity, and mass ratio) are a direct outcome of the formation process and the subsequent evolution of the systems. Multiple systems consisting of at least one substellar object are particularly important since, under the reasonable assumption of common formation, the main sequence components of the system also provide constraints on the age and metallicity of the companion, two parameters that are otherwise challenging to infer (e.g. Faherty et al. 2010; Pinfield et al. 2012; Burningham et al. 2013; Deacon et al. 2017; Chinchilla et al. 2020; Zhang et al. 2021a).

In particular, the frequency with which the lowest-mass stars and brown dwarfs exist as companions to higher mass stars, especially FGKM stars, is a crucial piece of evidence to distinguish between competing formation scenarios. Kroupa et al. (2013) argue that if stars and brown dwarfs formed via the same process, both sets ought to follow the same binary pairing rules. They note that simulations that assume common origin (Kroupa et al. 2001, 2003) overpredict the incidence of wide brown dwarf binaries and star plus brown dwarf pairs. Kroupa et al. (2013) therefore conclude that brown dwarfs form in a fundamentally different way from stars. On the other hand, Chabrier (2002, 2005) maintain that the low-mass end of the mass function is just a continuation of that seen at higher masses. Chabrier et al. (2014) argue that the paucity of wide brown dwarf companions is explained as the result of disintegration of such weakly bound systems through dynamical interactions. They thus conclude that brown dwarfs form the same way that higher mass stars do.

Observational constraints on the above theories are difficult to obtain, given the inherent difficulty of identifying faint, red companions against the dense backdrop of reddened distant stars and high-redshift quasars. Although Gl 229 B, one of the very first brown dwarfs discovered, is a wide T7 companion to an M1V star (Nakajima et al. 1995), late-T dwarfs as wide companions to M dwarfs remain rare, with ∼20 known to date (see Faherty et al. 2020; Zhang et al. 2021a; Kirkpatrick et al. 2023). Two recently released catalogs offer an unprecedented opportunity to identify more of these elusive M+T systems.

Released on 2022 June 13, *Gaia* DR3 provides exquisite astrometry for 1.812 billion objects, and represents a significant improvement over DR2 especially in terms of reliability of the nearby stars (Lindegren et al. 2021; Gaia Collaboration et al. 2021). *Gaia*, however, observes the sky at optical wavelengths and, therefore, cold brown dwarfs are almost completely invisible to it (Smart et al. 2019).

The CatWISE2020 Catalog (Eisenhardt et al. 2020; Marocco et al. 2021) complements *Gaia* DR3 by providing astrometry for 1.890 billion objects observed by NASA's *Wide-field Infrared Survey Explorer* (*WISE*; Wright et al. 2010) and by NEOWISE (the reactivated *WISE* satellite; Mainzer et al. 2014). Scanning the sky at 3.4 and 4.6 $\mu$m, *WISE* has the ability to detect the coldest substellar objects in the Solar neighborhood (e.g. Luhman 2014; Marocco et al. 2019; Bardalez Gagliuffi et al. 2020).

By combining *Gaia* DR3 with the CatWISE2020 Catalog, we have discovered 13 new nearby systems with T dwarfs as common proper motion companions in wide orbits around M stars, an increase of ∼60% over the known population.

In this paper, we begin by describing our target selection methodology in Section 2. In Section 3 we present our *Spitzer* photometric observations. In Section 4 we derive improved astrometric measurements for the newly-discovered T dwarfs, and use these measurements in Section 5 to assess the companionship probability of the binary pairs. Section 6 describes our spectroscopic follow-up for the new Ms and Ts. In Section 7 we use spectroscopic indicators, kinematics, and light curves for the M primaries to constrain the ages of the systems. Section 8 describes the individual systems in more detail, while Section 9 puts them in context with the rest of the low-mass companions population. Finally, Section 10 summarizes our findings.

## 2. TARGET SELECTION

### 2.1. *Identification of cold brown dwarf candidates*

Cold brown dwarf candidates were initially selected using *WISE* data following three complementary methods:

- Photometry, proper motion, and quality cuts applied to the CatWISE2020 Catalog.
- A machine-learning-based classifier trained on known cold brown dwarfs from the literature and applied to the CatWISE2020 Catalog.

- Visual inspection of unWISE epochal coadds (Meisner et al. 2019) by citizen scientists, through the "Backyard Worlds: Planet 9" collaboration (hereafter BYW).

The first two methods are described in detail in Meisner et al. (2020a) and Marocco et al. (2019), and the last method is described in Kuchner et al. (2017) and Meisner et al. (2020b). Here we provide only summaries of the procedures.

The photometry- and motion-based search is a combination of several different and complementary approaches, aimed at identifying an overall unbiased sample of cold brown dwarfs. A full listing of all different search criteria employed is given in Meisner et al. (2020a), and here we only provide a global view of their scope. Searches focused primarily on red, fast-moving sources, where the term "red" was implemented either via the color cut W1–W2≥1.5 mag or by selecting objects undetected in W1. The "fast-moving" term was implemented either through proper-motion selection with an emphasis on significance of motion (defined as $Q = e^{-\chi^2_{\rm motion}/2}$, where $\chi^2_{\rm motion} = (\mu_\alpha/\sigma_{\mu\alpha})^2 + (\mu_\delta/\sigma_{\mu\delta})^2$) or through reduced-proper-motion selection (defined as $H_{W2} = W2 + 5\log\mu_{\rm tot} + 5$, where $\mu_{\rm tot}$ is in arcsec yr$^{-1}$; Jones 1972). Artifact flags are used to remove spurious sources, or those with photometry badly contaminated by diffraction spikes, bright stars halos, latents or optical ghosts. Visual inspection of the candidates was performed using WISEView (Caselden et al. 2018) to remove sources with contaminated photometry or erroneous motion measurements.

The machine-learning-based search was conducted using the Python package *XGBoost* (Chen & Guestrin 2016), which implements machine-learning algorithms under the gradient boosting framework. The *XGBoost* classifier was trained on confirmed objects from the literature as the "positive" class, and a randomly selected sample of sources from the CatWISE2020 Catalog as the "negative" class. Sample weights and hyperparameters were chosen to minimize the classification error rate, which is defined as $n_{\rm wrong}/n_{\rm tot}$, where $n_{\rm wrong}$ is the number of misclassified objects, and $n_{\rm tot}$ is the total number of classified objects (see Tan 2018, Chapter 4.2). Once trained, the classifier was run on the entire CatWISE2020 Catalog to select objects with the highest probability of belonging to the "positive" class (i.e. of being cold brown dwarfs). This initial sample was visually inspected using the aforementioned WISEView program to remove misclassified objects. Further details on this selection method, as well as its overall yield, are presented in Marocco et al. (2019), Meisner et al. (2020a), and Kota et al. (2022).

BYW uses the Zooniverse web portal[1] to present citizen scientists with animated "flipbooks", each showing a $\sim 10' \times 10'$ patch of sky. These flipbooks are generated from the unWISE time-resolved coadds, which are combined into color-composite difference images. While stationary sources self-subtract, fast-moving, cold brown dwarfs appear as orange "dipoles". Citizen scientists are asked to flag any moving source in the flipbook. Citizen scientists can also submit their independent discoveries, obtained with methods of their choice (e.g. cross-matching of catalogs), directly to the BYW Core Science Team[2]. All sources submitted by the citizen scientists are vetted by the professional astronomers and, if found to be promising new candidates, followed-up to confirm/refute their nature.

## 2.2. *Identification of M dwarf primaries*

Most primaries were readily identified by eye by the citizen scientists during inspection of the $\sim 10' \times 10'$ flipbooks. Others were identified by our team of professional astronomers as part of the visual inspection of the cold brown dwarf candidates with WISEView, since the program blinks, by default, $2' \times 2'$ cutouts of the unWISE coadds, centered around the cold brown dwarf candidate under examination. Finally, we also identified primaries to the brown dwarf candidates by cross-matching the list of discoveries with *Gaia* DR2 using a $10'$ matching radius and requiring that the proper motion of the primary and putative companions agree at the $< 3\sigma$ level, i.e.

$$\left| \left(\sqrt{\mu_\alpha^{*\,2} + \mu_\delta^2}\right)_{\rm p} - \left(\sqrt{\mu_\alpha^{*\,2} + \mu_\delta^2}\right)_{\rm c} \right| < 3 \times \sqrt{\sigma^2_{\mu\alpha^*,\rm p} + \sigma^2_{\mu\delta,\rm p} + \sigma^2_{\mu\alpha^*,\rm c} + \sigma^2_{\mu\delta,\rm c}} \quad (1)$$

where the subscript "p" indicates measurements for the primary and the subscript "c" indicates measurements for the companion. Candidate pairs identified this way were then visually inspected and checked for consistency between the measured *Gaia* distance to the primary and the estimated photometric distance to the companion. Although the initial selection was done using *Gaia* DR2, when *Gaia* DR3 was released we updated the astrometry for our primaries to the newest values. The full list of 13 systems is presented in Table 1, and finder charts for all systems are shown in Appendix A.

[1] http://www.backyardworlds.org
[2] https://www.zooniverse.org/projects/marckuchner/backyard-worlds-planet-9/about/team

### 2.3. *Naming convention*

For systems where the primary has an entry in the SIMBAD astronomical database (Wenger et al. 2000), we use the SIMBAD name of the primary for both components and append an “A” to the name for the primary and a “B” to the name for the companion (e.g. L 26-16 A and L 26-16 B). For systems where the primary does not have an entry in SIMBAD, we use the CatWISE2020 *source_name* of the primary for both components and append an “A” to the name for the primary and a “B” to the name for the companion (e.g. CWISE J054129.32-745021.5 A and CWISE J054129.32-745021.5 B). The only exception to this convention is the system consisting of L 34-26 and WISEP J075108.79-763449.6, which is the only system where both the primary and the companion have a SIMBAD entry. In this case we decided to retain the SIMBAD name for both objects. We inspected the Washington Visual Double Star Catalogue (WDS; Mason et al. 2001) to ensure that our chosen names did not conflict with names already assigned therein.

**Table 1**. The new M+T systems. Coordinates for the M dwarfs are from *Gaia* DR3 (ICRS), while those for the T dwarfs are from CatWISE2020 (J2000). The angular separation ($\rho$) and position angle (P.A.) are computed at the CatWISE2020 epoch (2015.405) by proper-moving the M dwarf coordinates to this epoch using their *Gaia* DR3 astrometry. The position angle is measured East of North. The projected separation ($s$) assumes the system is at the distance of the M dwarf (listed in Table 4). Spectral types in parenthesis are based on photometry, otherwise they are based on spectroscopy (see Sections 6 and 8 for details). The last column lists, in alphabetical order, the co-discoverers of the system.

| ID | Short ID | R.A. | Decl. | $\rho$ | $s$ | P.A. | Sp. T. | Ref. | Discoverers |
|---|---|---|---|---|---|---|---|---|---|
| | | (hh:mm:ss.ss) | (dd:mm:ss.s) | (arcsec) | (au) | (deg) | | | |
| L 26-16 A | 0003A | 00:03:18.46 | −75:33:21.0 | 50.74 | 2199 | 282.45 | (M0 V) | 1 | AS, CW, DC, LG, MT, SG, WP |
| L 26-16 B | 0003B | 00:03:06.43 | −75:33:09.8 | | | | T4 | 1 | |
| 2MASS J00103250+1715490 A | 0010A | 00:10:32.50 | +17:15:49.0 | 18.09 | 739 | 316.42 | M8 V | 2 | AS, LG, SG |
| 2MASS J00103250+1715490 B | 0010B | 00:10:31.41 | +17:16:01.3 | | | | T5.5 | 1 | |
| UCAC3 52-1038 A | 0031A | 00:31:23.77 | −64:13:59.2 | 215.64 | 7139 | 45.31 | M2 V | 1 | AR, AS, CW, FK, LG, SG, ZW |
| UCAC3 52-1038 B | 0031B | 00:31:47.75 | −64:11:23.5 | | | | (T6) | 1 | |
| LP 712-16 A | 0312A | 03:12:14.52 | −08:45:45.8 | 16.93 | 623 | 242.46 | M4 V | 1 | CT, LG, SG, TB |
| LP 712-16 B | 0312B | 03:12:13.61 | −08:45:58.8 | | | | T6 | 1 | |
| UCAC3 40-6918 A | 0328A | 03:28:08.35 | −70:01:40.7 | 71.35 | 1549 | 5.02 | M3 V | 1 | AS, CW, LG, SG, TB |
| UCAC3 40-6918 B | 0328B | 03:28:09.72 | −70:00:26.8 | | | | (T8) | 1 | |
| CWISE J054129.32−745021.5 A | 0541A | 05:41:29.33 | −74:50:21.5 | 84.44 | 6576 | 264.41 | M4 V | 1 | AR, AS, MK, WP |
| CWISE J054129.32−745021.5 B | 0541B | 05:41:07.90 | −74:50:29.6 | | | | (T8) | 1 | |
| 2MASS J05581644−4501559 A | 0558A | 05:58:16.44 | −45:01:56.0 | 38.67 | 1043 | 174.73 | M4 V | 1 | AS, CW, DC, JW |
| 2MASS J05581644−4501559 B | 0558B | 05:58:16.68 | −45:02:33.5 | | | | T8 | 1 | |
| L 34-26 | 0749A | 07:49:12.68 | −76:42:06.7 | 596.77 | 6499 | 42.80 | M3 V | 3 | AR, CW, HH |
| WISEP J075108.79-763449.6 | 0749B | 07:51:08.71 | −76:34:50.1 | | | | T9 | 4 | |
| SCR J0959-3007 A | 0959A | 09:59:00.57 | −30:07:44.2 | 38.09 | 1165 | 310.67 | M5 V | 1 | AR, LG, SG, WP |
| SCR J0959-3007 B | 0959B | 09:58:58.01 | −30:07:18.1 | | | | (T5) | 1 | |
| UCAC4 307-069397 A | 1300A | 13:00:02.01 | −28:43:29.5 | 21.88 | 601 | 186.07 | M4 V | 1 | AS, LG, MT, SG, TB |
| UCAC4 307-069397 B | 1300B | 13:00:02.24 | −28:43:55.3 | | | | T6 | 1 | |
| LP 270-10 A | 1353A | 13:53:46.54 | +38:04:23.1 | 50.85 | 1750 | 199.06 | M2 V | 1 | AS, CW, DC, LG, MK, SG, TB |
| LP 270-10 B | 1353B | 13:53:45.23 | +38:03:32.1 | | | | T6.5 | 1 | |
| LP 81-30 A | 1416A | 14:16:11.37 | +23:23:29.2 | 55.29 | 1512 | 215.90 | M2 V | 1 | AS, CW, LG, MT, NS, SG, TB |
| LP 81-30 B | 1416B | 14:16:08.91 | +23:22:42.0 | | | | T7 | 1 | |
| G 135-35 A | 1417A | 14:17:39.95 | +20:56:28.9 | 151.71 | 4192 | 213.46 | (M3V) | 1 | AR, GC |
| G 135-35 B | 1417B | 14:17:33.98 | +20:54:22.3 | | | | (T3) | 1 | |

Note—Spectral type references: 1 - this paper; 2 - Gizis et al. (2000); 3 - Torres et al. (2006); 4 - Kirkpatrick et al. (2011). Discoverers code: AR - Austin Rothermich; AS - Arttu Sainio; CT - Christopher Tanner; CW - the CatWISE team; DC - Dan Caselden; FK - Frank Kiwy; GC - Giovanni Colombo; HH - Hiro Higashimura; JW - Jim Walla; LG - Léopold Gramaize; MK - Martin Kabatnik; MT - Melina Thévenot; NS - Nikolaj Stevnbak Andersen; SG - Samuel Goodman; TB - Thomas Bickle; WP - William Pendrill; ZW - Zbigniew Wedracki.

## 3. PHOTOMETRY

*Spitzer* photometric observations were taken for 5 T dwarfs as part of program 14034 (PI Meisner). Seven exposures of 30 s were taken in band ch1 (3.6$\mu$m) and ch2 (4.5$\mu$m), and these exposures were dithered using a random dither pattern of medium scale. The number of individual exposures was chosen so that we would obtain a 5$\sigma$ ch1 detection at ch1–ch2 = 2.75 mag.

Data reduction was performed using MOPEX (Makovoz et al. 2006). The data reduction is described in details in Kirkpatrick et al. (2019, Section 5.1) and Marocco et al. (2019, Section 4). Briefly, we performed both aperture and point response function (PRF) photometry using the *Spitzer* warm PRFs built by Jim Ingalls (see Kirkpatrick et al. 2019). Raw fluxes were converted to magnitudes using the correction factors and zero-points listed in the IRAC Instrument Handbook[3]. The difference between aperture and PRF photometry for our targets is negligible, so in the remainder of this paper we will use the PRF magnitudes, which are given in Table 2.

Additionally, we searched available large-area surveys to gather additional photometry for the Ms and Ts in our systems. We searched The Two Micron All Sky Survey (2MASS, Skrutskie et al. 2006), the United Kingdom Infrared Telescope (UKIRT) Infrared Deep Sky Survey (UKIDSS, Lawrence et al. 2007), the UKIRT Hemisphere Survey (UHS, Dye et al. 2018), and the Visible and Infrared Survey Telescope for Astronomy (VISTA) Hemisphere Survey (VHS, McMahon et al. 2013). We searched for initial matches using the CatWIE2020 coordinates and a generous search radius of 6″ to take into account the large epoch difference between the surveys considered and CatWISE2020. Subsequently, we visually inspected all matches to remove spurious counterparts. The photometry found is given in Tables 2 and 3.

## 4. ASTROMETRY

Astrometric information is crucial for the discovery of co-moving pairs and the assessment of their companionship. All primaries within our sample are well detected by *Gaia*, which provides excellent parallax and proper motion measurements (see, however, our discussions of 0541A in Section 8.6). None of the companions identified here is detected by *Gaia* given their cold temperatures. As discussed above (Section 2.1), our initial source of T dwarf proper motion measurements was the CatWISE2020 catalog. For several objects, however, their proximity to the very bright primary and the large full width at half maximum of the *WISE* point spread function (∼6″) mean that those measurements are prone to systematic uncertainties due to blending and/or contamination by the primary. External, high-signal-to-noise-ratio data obtained with instruments with a narrower point spread function can help us constrain the proper motion measurements further. We used the aforementioned *Spitzer*/IRAC observations, as well as UKIDSS, UHS, and VHS archival data. We combined these observations with the positions measured from the unWISE time-resolved coadds (see Meisner et al. 2018 for details on how the time-resolved unWISE coadds are constructed, and Meisner et al. 2023 for details on how source centroids are measured). The individual positions and their epochs are given in Appendix B.

To remove systematic offsets between the data sets, we re-registered the individual epochs using *Gaia* DR3 as follows. For the *Spitzer* data, we measured the position of all stars in the images using MOPEX/APEX (Makovoz & Marleau 2005), following the same procedure described in Kirkpatrick et al. (2021a). For the UKIDSS, UHS, and VHS data, we retrieved all catalogued sources within 10′ of our T dwarfs.

For each target, we cross-matched the resulting list of stars at each epoch to establish a reference set for astrometric recalibration. The cross-match used a radius of 5″, and only retained unsaturated stars that appear in all epochs, and measured with S/N > 10 in the respective bands. The typical reference set for each target consisted of ∼100 stars. We then cross-matched these reference sets with *Gaia* DR3, using again a 5″ radius. When performing this cross-match, the *Gaia* positions are first recomputed at the epoch of the external observation using the *Gaia* proper motion and parallax. To exclude possible unresolved binaries from the reference set, we selected stars with re-normalized unit weight error (hereafter *ruwe*) less than 1.4, as recommended by Lindegren et al. (2018).

Next, for a given reference star observed at a given epoch $t$, we define the following 6-parameter transformation:

$$\alpha_{G,t} = A_0 + A_1\alpha_t + A_2\delta_t \quad (2)$$

$$\delta_{G,t} = B_0 + B_1\alpha_t + B_2\delta_t \quad (3)$$

where $\alpha_{G,t}, \delta_{G,t}$ are the coordinates measured by *Gaia* and propagated to the epoch $t$, $\alpha_t, \delta_t$ are the coordinates measured at the epoch $t$, and the $A_i, B_i$ are the coefficients of the transformation. This accounts for offset,

[3] https://irsa.ipac.caltech.edu/data/SPITZER/docs/irac/iracinstrumenthandbook/

**Table 2.** Photometry for the T dwarf companions. The W1 and W2 magnitudes are the *w1mpro_pm* and *w2mpro_pm* from CatWISE2020. *Spitzer* photometry is from our dedicated observing campaign (see Section 3). J and $K_s$ magnitudes are in the MKO system. The J magnitude for WISEP J075108.79-763449.6 is from Kirkpatrick et al. (2011, K11). The only source with H-band photometry is WISEP J075108.79-763449.6, which has H = 19.68±0.13 (Kirkpatrick et al. 2019).

| Short ID | ch1 | ch2 | W1 | W2 | $J$ | $J$ src | $K_s$ | $K_s$ src |
|---|---|---|---|---|---|---|---|---|
| | (mag) | (mag) | (mag) | (mag) | mag | | (mag) | |
| 0003B | 15.962±0.029 | 15.497±0.024 | 16.704±0.032 | 15.620±0.035 | 17.487±0.034 | VHS | 17.471±0.144 | VHS |
| 0010B | ... | ... | 17.523±0.080 | 15.919±0.060 | 17.528±0.034 | UHS | ... | ... |
| 0031B | ... | ... | 17.332±0.053 | 15.305±0.027 | 17.490±0.020 | VHS | 17.517±0.101 | VHS |
| 0312B | ... | ... | 16.832±0.050 | 15.423±0.041 | ... | ... | ... | ... |
| 0328B | 18.319±0.151 | 15.926±0.024 | 19.100±0.189 | 16.100±0.043 | 19.582±0.173 | VHS | ... | ... |
| 0541B | ... | ... | 18.963±0.174 | 16.077±0.042 | ... | ... | ... | ... |
| 0558B | 17.634±0.092 | 15.855±0.026 | 18.455±0.114 | 15.876±0.039 | 19.501±0.089 | VHS | ... | ... |
| 0749B | ... | ... | 17.080±0.036 | 14.610±0.015 | 19.34±0.05 | K11 | ... | ... |
| 0959B | ... | ... | 17.392±0.070 | 15.697±0.047 | 18.239±0.055 | VHS | ... | ... |
| 1300B | ... | ... | 16.810±0.052 | 14.901±0.027 | 17.535±0.028 | VHS | 17.959±0.196 | VHS |
| 1353B | 17.084±0.064 | 15.681±0.026 | 17.987±0.103 | 15.796±0.046 | 18.515±0.080 | UHS | ... | ... |
| 1416B | 16.678±0.047 | 15.247±0.022 | 17.868±0.098 | 15.413±0.034 | 17.651±0.030 | UHS | ... | ... |
| 1417B | ... | ... | 17.296±0.059 | 16.357±0.078 | ... | ... | ... | ... |

**Table 3.** Photometry for the M dwarf primaries. $G$, $G_{\rm BP}$ and $G_{RP}$ are from *Gaia* DR3. $J$, $H$, and $K_{\rm s}$ are from 2MASS. W1 and W2 are the *w1mpro_pm* and *w2mpro_pm* from CatWISE2020.

| Short ID | $G$ | $G_{\rm BP}$ | $G_{\rm RP}$ | $J$ | $H$ | $K_{\rm s}$ | W1 | W2 |
|---|---|---|---|---|---|---|---|---|
| | (mag) | (mag) | mag | (mag) | (mag) | (mag) | (mag) | (mag) |
| 0003A | 11.0844±0.0028 | 11.9248±0.0029 | 10.1864±0.0038 | 9.097±0.021 | 8.415±0.051 | 8.213±0.018 | 8.238±0.012 | 8.197±0.008 |
| 0010A | 18.2442±0.0035 | 21.25±0.14 | 16.6867±0.0073 | 13.895±0.028 | 13.211±0.026 | 12.799±0.027 | 12.624±0.012 | 12.405±0.009 |
| 0031A | 12.2090±0.0028 | 13.3462±0.0030 | 11.1512±0.0038 | 9.812±0.026 | 9.248±0.023 | 8.991±0.019 | 8.868±0.012 | 8.729±0.007 |
| 0312A | 13.7810±0.0028 | 15.2500±0.0031 | 12.6009±0.0039 | 10.976±0.026 | 10.458±0.024 | 10.179±0.025 | 10.042±0.013 | 9.882±0.009 |
| 0328A | 11.9962±0.0028 | 13.2875±0.0030 | 10.8741±0.0038 | 9.413±0.024 | 8.812±0.029 | 8.585±0.025 | 8.466±0.013 | 8.289±0.010 |
| 0541A | 14.9937±0.0038 | 16.3171±0.0041 | 13.6176±0.0039 | 11.979±0.025 | 11.399±0.026 | 11.125±0.025 | 10.970±0.013 | 10.799±0.013 |
| 0558A | 13.4906±0.0028 | 15.1686±0.0039 | 12.2493±0.0041 | 10.480±0.026 | 9.918±0.026 | 9.624±0.023 | 9.490±0.012 | 9.294±0.009 |
| 0749A | 10.1621±0.0030 | 11.5583±0.0053 | 8.9939±0.0044 | 7.406±0.021 | 6.862±0.031 | 6.579±0.018 | 7.136±0.021 | 6.448±0.010 |
| 0959A | 14.8034±0.0029 | 16.7871±0.0069 | 13.4909±0.0042 | 11.466±0.024 | 10.886±0.023 | 10.600±0.019 | 10.453±0.012 | 10.262±0.008 |
| 1300A | 12.5693±0.0028 | 13.9023±0.0036 | 11.4243±0.0038 | 9.917±0.023 | 9.360±0.022 | 9.104±0.019 | 9.002±0.012 | 8.817±0.008 |
| 1353A | 12.7966±0.0028 | 13.9854±0.0029 | 11.7129±0.0038 | 10.282±0.020 | 9.739±0.019 | 9.493±0.017 | 9.384±0.014 | 9.226±0.009 |
| 1416A | 11.0771±0.0028 | 12.1409±0.0030 | 10.0520±0.0038 | 8.713±0.020 | 8.167±0.027 | 7.928±0.024 | 7.869±0.014 | 7.733±0.008 |
| 1417A | 12.3898±0.0028 | 13.6542±0.0030 | 11.2733±0.0038 | 9.819±0.021 | 9.290±0.020 | 9.062±0.017 | 9.214±0.011 | 9.030±0.009 |

rotation, and scale between the measured coordinates and the *Gaia* ones.

Finally, we determined the 6 parameters at each epoch via $\chi^2$ fitting. The transformation was then applied to the coordinates of our target, and the resulting reregistered coordinates were used to measure the $\mu_\alpha, \mu_\delta$ components of the proper motion through a linear fit. The resulting proper motions are listed in Table 4.

For four of the thirteen T dwarfs we did not recompute proper motions. For 0010B and WISEP J075108.79-763449.6 the literature astrometry is of very high quality, so the procedure described above would not improve upon it. For 0541B and 1417B there is no external data that would allow us to improve upon the CatWISE2020 values.

## 5. ASSESSMENT OF COMPANIONSHIP PROBABILITY

A fundamental step in the discovery of new, widely separated binaries is the assessment of the probability

**Table 4.** Astrometry for the new binary systems. Literature proper motions and distances for the M dwarfs are from *Gaia* DR3. Literature proper motions for all the T dwarfs except 0010B and 0749B are from CatWISE2020, while their distances are photometric estimates. 0010B has a measured proper motion from Schneider et al. (2023). WISEP J075108.79-763449.6 has a measured proper motion and astrometric distance from Kirkpatrick et al. (2019). The last column shows the companionship probability computed in Section 5.

| | Literature | | This paper | | | |
|---|---|---|---|---|---|---|
| Short ID | $\mu_\alpha \cos\delta$ (mas yr$^{-1}$) | $\mu_\delta$ (mas yr$^{-1}$) | $\mu_\alpha \cos\delta$ (mas yr$^{-1}$) | $\mu_\delta$ (mas yr$^{-1}$) | d (pc) | Comp. Prob. (%) |
| 0003A | 303.214±0.052 | 23.696±0.044 | ... | ... | 43.470±0.048 | 99.9 |
| 0003B | 283±43 | 70±38 | 298±49 | 70±45 | 35 | |
| 0010A | −184.54 ± 0.41 | −42.24 ± 0.29 | ... | ... | 40.501736 ± 0.37 | 100.0 |
| 0010B | −182.9 ± 12.8 | −39.2 ± 12.7 | ... | ... | 42 | |
| 0031A | 227.593±0.038 | 269.212±0.036 | ... | ... | 33.182±0.026 | 100.0 |
| 0031B | 181±48 | 270±43 | 236±37 | 271±33 | 30 | |
| 0312A | 98.105±0.086 | −323.254 ± 0.086 | ... | ... | 36.72±0.068 | 0.0[a] |
| 0312B | 610±39 | −153±39 | 252±54 | −150±51 | 36 | |
| 0328A | 87.342±0.043 | 286.906±0.046 | ... | ... | 21.722±0.011 | 99.8 |
| 0328B | 132±78 | 306±71 | 110±65 | 344±62 | 20 | |
| 0541A | 19.29±0.33 | 119.31±0.40 | ... | ... | $77.9^{+1.8}_{-1.7}$ | 0.0[b] |
| 0541B | −38±51 | 227±57 | ... | ... | 34 | |
| 0558A | −67.024±0.066 | 61.849±0.071 | ... | ... | 26.969±0.027 | 99.9 |
| 0558B | −133±66 | 76±76 | −97±53 | 109±54 | 29 | |
| 0749A | −101.997±0.061 | −193.265±0.068 | ... | ... | 10.8898±0.0041 | 100.0 |
| 0749B | −104.8±2.8 | −189.7±4.5 | ... | ... | $10.21^{+0.75}_{-0.65}$ | |
| 0959A | −304.27±0.14 | 68.64±0.13 | ... | ... | 30.76±0.10 | 99.5 |
| 0959B | −323.4±68.9 | 16.8±75.4 | −293±66 | −4±71 | 42 | |
| 1300A | 330.994±0.094 | −281.921±0.071 | ... | ... | 27.438±0.034 | 100.0 |
| 1300B | 319±30 | −13±33 | 338±59 | −297±49 | 28 | |
| 1353A | 74.846±0.028 | −209.731±0.041 | ... | ... | 34.385±0.034 | 99.9 |
| 1353B | 183±78 | −199±88 | 85±64 | −222±71 | 32 | |
| 1416A | −110.175±0.16 | −170.56±0.17 | ... | ... | 27.452±0.065 | 100.0 |
| 1416B | −249±57 | −127±63 | −244±46 | −160±47 | 26 | |
| 1417A | −136.325±0.020 | −614.950±0.021 | ... | ... | 27.631±0.019 | 0.0[c] |
| 1417B | −37±59 | −197±61 | ... | ... | 43 | |

Note—[a] This probability is computed using the *Gaia* proper motion and parallax for the primary and the CatWISE proper motion and photometric distance for the companion. For further assessment on the companionship probability of this pair see Section 8.4. [b] This probability is computed using the *Gaia* proper motion and parallax for the primary and the CatWISE proper motion and photometric distance for the companion. For further assessment on the companionship probability of this pair see Section 8.6. [c] This probability is computed using the *Gaia* proper motion and parallax for the primary and the CatWISE proper motion and photometric distance for the companion. For further assessment on the companionship probability of this pair see Section 8.13.

that the system in question is merely the result of a chance alignment between two unrelated sources. This is particularly important in our case since we do not have, in most cases, direct measurements of the distance to the T dwarf components of the putative systems and, in a few cases, the proper motion measurements themselves are highly uncertain.

To estimate the probability that each pair presented here forms a physically bound system, we used Co-Mover (Gagné et al. 2021b). This program uses the coordinates, proper motions, and optionally parallaxes and radial velocities of the two components of the system. It then builds a multivariate 6-dimension Gaussian model from the kinematic information of the primary (the Galactic XYZ coordinates and the UVW components of the Galactic velocity), or a series of models if some of the kinematic information is missing. The program then compares the observed kinematics of the putative companion to this model as well as to a 10-component multivariate Gaussian model for field stars[4]. The comparison is done using Bayes' theorem, and the code returns the probability that the two objects are related.

We used the *Gaia* DR3 coordinates, proper motions, and parallaxes for the M dwarf primaries, and the CatWISE2020 coordinates, our measured proper motions (see Section 4) and estimated photometric distances for the T dwarfs. The photometric distance estimates are obtained using either the CatWISE2020 or, if available, the *Spitzer* photometry and the spectral types listed in Table 1 (see Section 6) with the type-to-absolute-magnitude relations from Kirkpatrick et al. (2021a).

The companionship probabilities are presented in the rightmost column of Table 4. Pairs with very low companionship probability are described in further detail in their respective subsections in Section 8.

## 6. SPECTROSCOPIC FOLLOW-UP

### 6.1. *Lick/Kast*

Optical spectra of 0312A and 1300A were obtained with the Kast spectrograph (Miller & Stone 1994) on the Lick 3m Shane Telescope[5]. 0312A was observed on UT 2021 January 10, while 1300A was observed on UT 2021 May 14. Observations were conducted with the 2″ slit and 600/7500 red grating, providing resolution $\lambda/\Delta\lambda \approx 1800$ over the 6300–9000 Å wavelength range. For all targets we obtained 2 exposures of 600 seconds each. Data were reduced using the `kastredux` package[6] using default settings.

### 6.2. *Gemini/Flamingos-2*

0003B was observed with the Flamingos-2 instrument (Eikenberry et al. 2004) on Gemini South on UT 2019 June 19. The spectra were obtained with the 4-pixel wide (0.″72) long slit (4.′4) using the JH grism, which resulted in an average resolving power of ∼350. Thirty-two exposures of 120 seconds were obtained in a repeating ABBA pattern for a total exposure time of 64 minutes. The A3V star HIP 116234 was used for telluric corrections. We used the Gemini `IRAF` data reduction package[7] to process the spectra and followed the standard procedures outlined in the Flamingos-2 Longslit Tutorial[8].

### 6.3. *Keck/NIRES*

Near-infrared spectra for 0010A and B, and 1416B were obtained using the Near-Infrared Echellette Spectrometer[9] (NIRES, Wilson et al. 2004) on the Keck 2 telescope. 0010A and B were observed on UT 2019 December 19, while 1416B was observed on UT 2020 July 20. Data were reduced using a modified version of Spextool (Cushing et al. 2004) with standard settings. Reduction steps included spectral order rectification and pixel response calibration using dome flat-field lamp observations, wavelength calibration using OH emission lines in deep exposures, optimal extraction of point source spectra and combination of multiple exposures using a sigma-clipped weighted mean after removal of cosmic ray hits, and correction of telluric absorption and instrumental response calibration using spectra of A0 V stars observed after target observations at a similar airmass using the methodology of Vacca et al. (2003). Individual spectral orders were stitched together manually to account for inter-order flux scaling variations. The final data had median signal-to-noise ratios of 25–75 at 1.27 $\mu$m.

### 6.4. *Magellan/FIRE*

0312B, 0558B, and 1300B were observed with the Folded-port Infrared Echellete spectrograph (FIRE; Simcoe et al. 2013) at the 6.5 m Baade Magellan telescope. We used the high-throughput prism mode with a 0.″6 slit, which gives a resolving power ($\lambda/\Delta\lambda$) of $\sim$450 across the 0.8–2.45 $\mu$m range. 0312B, 0558B, and 1300B were observed on UT 2019 December 11, UT 2020 February 13, and UT 2020 February 12, respectively. Each target was observed using the sample-up-the-ramp mode and nodded along the slit. A0V stars were observed immediately after each science target for telluric correction purposes. For 0558B, we obtained twelve 126.8 second exposures, giving a total on-source exposure time of 1522 seconds. We obtained ten 84.5 second exposures for 1300B, resulting in a total on-source time of 845 seconds. Reductions were performed with a modified version of the `FIREHOSE` package (Gagné et al. 2015).

[4] The 10-component model is used to represent the non-Gaussian XYZUVW distribution of nearby field stars. This approximation is valid for stars within a few hundred parsecs of the Sun

[5] https://mthamilton.ucolick.org/techdocs/instruments/kast/

[6] https://github.com/aburgasser/kastredux

[7] https://www.gemini.edu/observing/phase-iii/reducing-data/gemini-iraf-data-reduction-software

[8] https://gemini-iraf-flamingos-2-cookbook.readthedocs.io/en/latest/Tutorial_Longslit.html

[9] https://www2.keck.hawaii.edu/inst/nires/

### 6.5. *SALT/RSS*

0031A, 0328A, 0558A, 0541A, and 0959A were observed with the Robert Stobie Spectrograph (RSS) on the Southern African Large Telescope (SALT) on UT 2021 December 25. The spectrograph was used in long slit mode using the PG0900 grating at an angle of 20°, which produces coverage over the ranges 6033–7028, 7079–8045, and 8091–9023 Å across the 3×1 mini-mosaic, delivering a resolution of $\sim$600 in the short-wavelength portion, increasing to $\sim$2000 in the long-wavelength portion. The spectra were reduced following the procedure described in Kirkpatrick et al. (2023).

### 6.6. *Spectral typing*

We assigned a spectral type to targets for which we had optical and/or near-infrared spectroscopy via visual matching to standard templates. We used the standard templates defined in Kirkpatrick et al. (2016, M0–M9), Kirkpatrick et al. (2010, L0–L9), Burgasser et al. (2006, T0–T8), and Cushing et al. (2011, T9–Y1). For the T dwarfs, we followed the prescriptions of Kirkpatrick et al. (2010), i.e. we selected the template that provided the best match to the $J$-band portion of the observed spectrum. The results from template matching are presented in the individual subsections of Section 8, and the assigned spectral types are listed in Table 1.

Objects that were not followed-up spectroscopically were typed using their photometry. For 0003A and 1417A, the only M primaries lacking spectroscopy, we used the *Virtual Observatory SED Analyzer*[10] (hereafter VOSA; Bayo et al. 2008) to gather available photometry from numerous surveys, spanning the UV to mid-IR range. We then used VOSA to fit the SED with the M dwarf templates from Kesseli et al. (2017). For T dwarfs without spectroscopy, we used the available *Spitzer* photometry or, lacking that, CatWISE2020 photometry and estimated their spectral type with the color-type relations from Kirkpatrick et al. (2021a). Spectral types determined using these two methods are listed in parentheses in Table 1.

Further details are given in the individual subsections of Section 8.

## 7. AGE DETERMINATION

The ages of our binary systems can be constrained using the M dwarf primaries and their measured properties. In this paper, we follow an approach similar to the one described in Schneider et al. (2021), who employed a combination of spectroscopy, time-resolved photometry, activity indicators, and kinematics to constrain the age of Ross 19A.

### 7.1. *TESS light curves*

The rotation rate of stars decreases with age due to angular momentum loss as a result of the interaction between the magnetic field and the stellar wind (Skumanich 1972). While this phenomenon is well understood for FGK stars (e.g. Barnes 2003), the picture is more complicated for M dwarfs. The study of large samples of M dwarfs have shown that gyrochronology is applicable to low-mass stars too, but at ages $\lesssim$400–700 Myr the intrinsic scatter in rotation rates (due to the spread in initial angular momentum) makes precise age determination for individual stars challenging (e.g. Rebull et al. 2016; Newton et al. 2018; Popinchalk et al. 2021). Nevertheless, measuring the rotation rate for M dwarfs can still give us an indication of their approximate age.

All M primaries except 0010A have been observed by the *Transiting Exoplanet Survey Satellite* (*TESS*; Ricker et al. 2015). We used both the 30-minute and, when available, the 2-minute light curves to measure the rotation periods of our M dwarfs.

The 30-minute light curves were extracted from the *TESS* full-frame images (hereafter FFIs) using the publicly available `Python` package *lightkurve* (Lightkurve Collaboration et al. 2018). For the 2-minute light curves, we retrieved the pipeline-produced light curves available on the *Barbara A. Mikulski Archive for Space Telescopes* (MAST). Rotation periods were identified using a box-least-square periodogram (Kovács et al. 2002), and the folded light curves were visually inspected to assess the reliability of the period determination and to identify any possible peculiarity. The periodic signal was then removed from the light curve using *wotan* (Hippke et al. 2019) and the periodogram recalculated to identify additional periodic signals. We detect clear variability and, therefore, measure rotation periods for 4 out of 12 M dwarf primaries observed by *TESS*. The measured periods are listed in Table 5.

[10] http://svo2.cab.inta-csic.es/theory/vosa/index.php

We can derive age constraints based on these rotation periods by comparing them to rotation periods of stars of similar spectral type and known age. Figure 1 shows the logarithm of the rotation period as a function of *Gaia* $G - G_{RP}$ color, with bluer and larger objects to the left, and smaller and redder objects to the right. We use *Gaia* $G - G_{RP}$ as Kiman et al. (2019) shows it most effective for describing M dwarf and the cool stars regime. In the background are three comparison populations, the youngest being the Pleiades at 120 Myr (Rebull et al. 2016), then Praespe at 650 Myr (Douglas et al. 2014, 2019), and finally a smattering of field stars from K2 (Popinchalk et al. 2021) and MEarth (Newton et al. 2016, 2018) that represent objects thought to be billions of years old.

The 4 objects for which a rotation period was measured are shown as stars. Given the large scatter observed in the reference populations, it is challenging to constrain the age of our targets precisely, but their rapid rotation implies youth of some kind (i.e. $\lesssim$1 Gyr).

For the remaining 8 M primaries, *TESS* does not reveal clear variability. Not observing variability in a light curve can be indicative of an old age for the object, as amplitude of stellar variability is thought to decrease with age (see Morris 2020 for a study of FGK stars). However, it could also be due to the observation window used (i.e. the rotation period could be much longer of the *TESS* baseline), or to a period of reduced stellar activity of the star, or even to the target being observed pole-on, which makes the rotation imperceptible. Therefore the lack of rotation-driven variability does not necessarily rule out a young age for any of these stars.

### 7.2. *Hα analysis*

The strength of the Hα line is known to be a good proxy for activity in M dwarfs, especially at young ages (Kiman et al. 2021, and references therein). We collected Hα measurements from the literature, and complemented them with our own measurements from follow-up spectroscopy. The equivalent width is defined as:

$$\mathrm{EW} = \int_{\lambda_1}^{\lambda_2} \left(1 - \frac{F_\lambda}{F_c}\right) d\lambda \quad (4)$$

where $F_\lambda$ is the flux density of the source spectrum and $F_c$ is the continuum flux density. The continuum for each spectrum was taken as the mean flux density across the 6500-6550 Å and 6575-6625 Å regions following West et al. (2011). The integral was evaluated over an 8 Å window centered at the peak of the Hα emission. In practice, since spectra are measured over discrete pixels, we summed the flux in all pixels that overlapped this window.

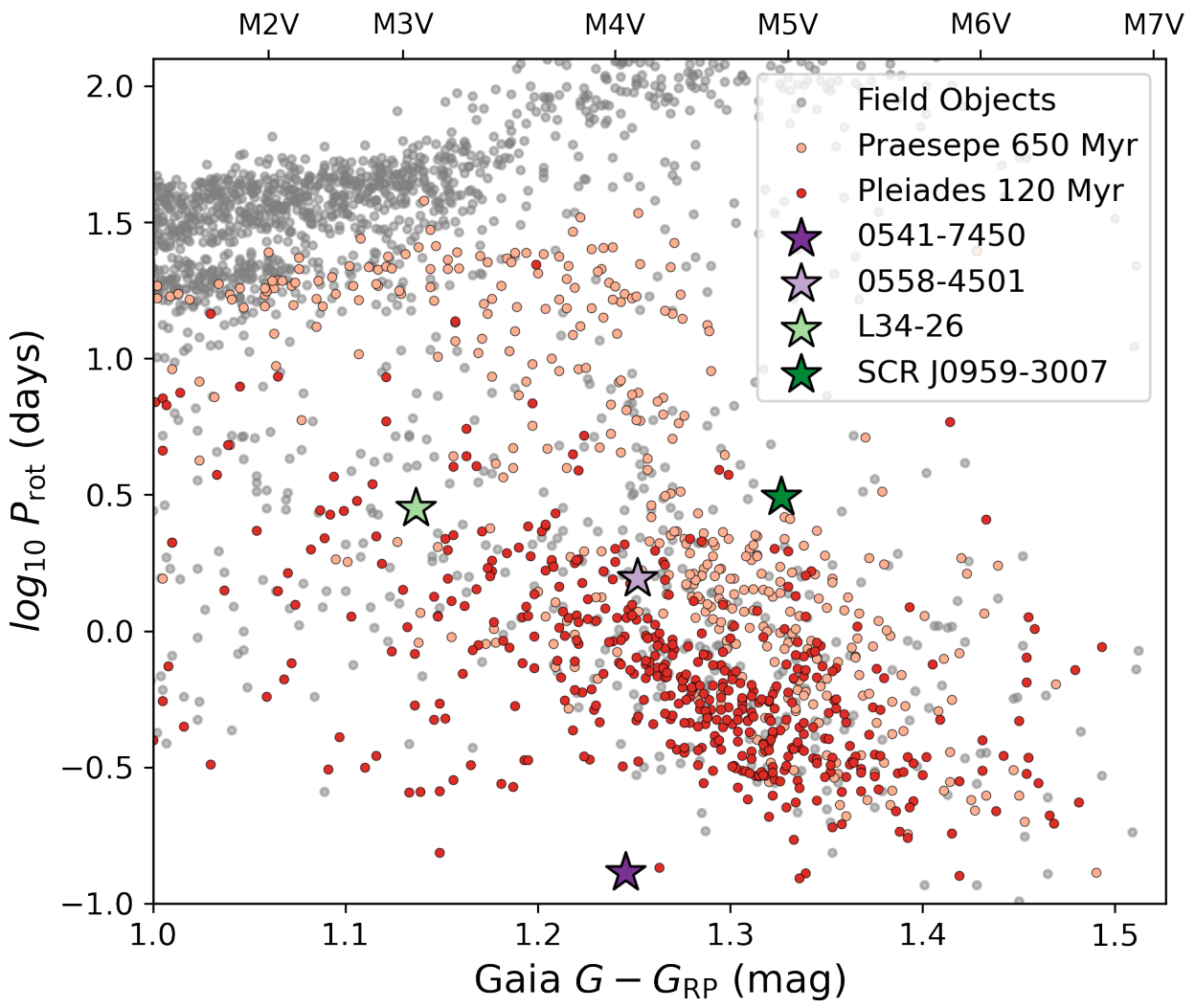

**Figure 1.** Rotation period as a function of *Gaia* $G - G_{RP}$ color for the four primaries for which we observe rotationally-induced variability. Overplotted for comparison are objects from two young clusters (the Pleiades – 120 Myr – and Praesepe – 650 Myr) as well as "old" objects from the field (i.e. older than 1 Gyr). An approximate spectral type scale is shown over the top of the plot.

To derive age constraints from these measurements, we used the broken-power-law age-activity relation from Kiman et al. (2021). We only applied the relation to objects whose $H\alpha$ equivalent width is above the activity threshold defined by Equation 1 of Kiman et al. (2021). The results for each system are presented in the individual subsections of Section 8 and summarized in Table 5.

### 7.3. *UV and X-ray emission*

Stellar activity is a well-established indicator of youth (e.g. Preibisch & Feigelson 2005) in late-type stars, because it is part of the rotation-age-activity relation. In turn, UV and X-ray emission are excellent indicators for the activity level of M dwarfs, as they probe chromospheric and coronal non-thermal processes. As such, UV and X-ray emission provide additional powerful tools to constrain the age of M dwarfs.

Shkolnik et al. (2011) and Rodriguez et al. (2013) used combinations of UV and IR photometry to identify potentially young M dwarfs in the Solar neighborhood. Shkolnik et al. (2011) used the ratio between the NUV flux and the J-band flux, while Rodriguez et al. (2013)

**Table 5.** Age determination for the 13 M dwarf primaries. For each object we list the rotation period determined from the *TESS* light curves and the corresponding age estimate ($P_{\rm rot}$, Age$_{\rm Prot}$), the equivalent width of the $H\alpha$ line and the implied age (EW $H\alpha$, Age$_{H\alpha}$), the age constrain from the kinematics (Age$_{\rm Kin}$), any age constraint from the literature (Age$_{\rm Lit}$), and, in the last column, the adopted age for the system.

| Short ID | $P_{\rm rot}$ | Age$_{\rm Prot}$ | EW $H\alpha$ | Age$_{H\alpha}$ | Age$_{\rm Kin}$ | Age$_{\rm Lit}$ | Adopted age |
|---|---|---|---|---|---|---|---|
| | (d) | (Gyr) | (Å) | (Gyr) | (Gyr) | (Gyr) | (Gyr) |
| 0003A | ... | ... | n/a | n/a | 3.4–10.6 | 1.8 | 1.8–10.6 |
| 0010A | n/a | n/a | n/a | n/a | 2.8–10.1 | ... | 2.8–10.1 |
| 0031A | ... | ... | ∼0 | >1 | 2.5–9.9 | ... | 2.5–9.9 |
| 0312A | ... | ... | 0.11±0.05 | >1 | 3.3–10.6 | ... | 3.3–10.6 |
| 0328A | ... | ... | ∼0 | >1 | 3.3–10.6 | 0.79 | 0.79–10.6 |
| 0541A | 0.13 | <0.12 | ∼0 | >1 | 2.9–10.3 | ... | 0.12–1 |
| 0558A | 1.56 | 0.12–0.65 | 2.32±0.05 | <1 | 2.4–9.6 | ... | 0.12–0.65 |
| 0749A | 2.83 | <0.65 | 2.4–8.0 | <1 | 2.3–9.0 | ... | <0.65 |
| 0959A | 3.10 | <1 | ∼0 | >1 | 3.0–10.3 | ... | <1 |
| 1300A | ... | ... | 0.19±0.05 | >1 | 3.3–10.5 | ... | 3.3–10.5 |
| 1353A | ... | ... | 0.12±0.05 | >1 | 2.6–9.5 | ... | 2.6–9.5 |
| 1416A | ... | ... | 0.17±0.05 | >1 | 2.7–9.9 | 3.1 | 2.7–9.9 |
| 1417A | ... | ... | n/a | n/a | 4.1–11.7 | ... | 4.1–11.7 |

used the NUV–W1 and J–W2 colors (see their Figure 1).

We collected UV measurements for the M dwarf primaries from the *Galaxy Evolution Explorer* (*GALEX*; Martin et al. 2003; Bianchi et al. 2017) All-Sky Imaging Survey (AIS). We cross-matched our list of primaries with the AIS using a 20″ radius, and retained only the nearest match for each M dwarf. We then visually inspected the AIS images, comparing them against higher angular resolution optical and NIR images (the *GALEX* FWHM is 4″) to rule out the presence of background sources that could contaminate the UV measurement, or lead to spurious measurements altogether. When an M dwarf was undetected in the AIS, we estimated an empirical $3\sigma$ limit on its FUV and NUV flux by querying the AIS in a radius of 30′ around the target, and taking the median flux for sources with S/N∼3 within that radius as the limit.

We used the same procedure to collect X-ray measurements from the *Röntgensatellit* (*ROSAT;* Truemper 1982) all-sky survey bright source catalogue (Voges et al. 1999). Visual inspection of the images is, in this case, crucial given the much lower angular resolution of the *ROSAT* images (∼1.8′/pix). We did not attempt to estimate a $3\sigma$ limit for X-ray non-detections.

We compared the UV and X-ray measurements (and limits) for our M dwarfs with the samples of young M dwarfs presented in Shkolnik et al. (2011) and Rodriguez et al. (2013) to derive qualitative limits on the age of our targets. The results are discussed in the subsections of Section 8.

### 7.4. *Kinematics*

Kinematic heating, i.e. the increase in the width of the velocity distribution of stars as a result of gravitational interaction with giant molecular clouds, has long been used as a way to estimate the age of a stellar population (Wielen 1977). Constraining the age of an individual star via the same method is much more challenging, and even more with incomplete kinematics information. In this paper, we attempt to constrain the age of our M dwarfs following three different methodologies. The first approach follows Schneider et al. (2021) and Burgasser & Mamajek (2017), as we describe in the following paragraphs.

We compared the kinematics of our sample with that of age-calibrated samples from the Spectroscopic Properties of Cool Stars survey (SPOCS, Valenti & Fischer 2005), the Geneva-Copenhagen Survey (GCS, Casagrande et al. 2011), Bensby et al. (2014), Brewer et al. (2016), and Luck (2017, 2018).

For each M dwarf primary, we computed its tangential velocity using its *Gaia* parallax and proper motion. If a radial velocity was available from the literature, we combined it with the tangential velocity to compute the total velocity of the M dwarf. We then selected stars from each reference sample with either tangential or total velocities within 15 km $s^{-1}$ of the corresponding velocity of the M dwarf (depending on what measurement is available), to ensure at least 10 comparison stars from each sample. A larger velocity range would smooth over age gradients across velocity space given that, for example, the transition between thin disk and thick disk stars happens over a range of ∼20 km $s^{-1}$ (e.g. Bensby et al. 2014). We then assumed a uniform probability distribution of age for each star in the above sub-samples between either the minimum and maximum ages provided (e.g., SPOCS) or between the 16% and 84% isochronal ages (e.g., GCS). We constructed a combined age probability distribution by combining the individual age probability distributions using a Monte Carlo approach, giving equal weight to each sample of stars. An example for the resulting combined age probability distribution function (PDF) is shown in Figure 2 for the M dwarf primary 0003A.

The resultant 16%–84% kinematic age ranges are listed in Table 5, and discussed in Section 8.

It is important to remember than any age bias in the reference populations used in this method will be imprinted in our kinematic age estimate. While combining different samples somewhat alleviate this issue, one major limit remains – all reference samples consist of stars older than ∼2 Gyr. While we can mathematically extrapolate the age PDF down to 0 Gyr, this extrapolation is not validated empirically, and therefore a) kinematic age estimates $\lesssim$ 2 Gyr are unreliable, and b) our kinematic age analysis is overall biased towards older ages, i.e. our method will tend to overestimate the age of young M dwarfs. These are, most likely, the reason for the discrepancies between kinematic ages and other age estimates listed in Table 5.

The second approach follows Bensby et al. (2003). Assuming that the distribution of UVW velocity for the three components of the Galaxy (i.e. thin disk, thick disk, and halo) are gaussian, the probability for an object to belong to each of the three components is calculated using equation 1 and 2 from Bensby et al. (2003). We took the $\sigma$ of the three distributions from Bensby et al. (2003). All but two of the M dwarfs for which we have complete kinematic information (i.e. proper motion, parallax, and radial velocity; see Table 4) have a probability >90% of belonging to the thin disk. 0328A has a probability of ∼14% of belonging to the thick disk and ∼86% of belonging to the thin disk. 1417A has a probability of 99.5% of belonging to the thick disk, and only 0.2% and 0.3% of belonging to the thin disk and halo, respectively.

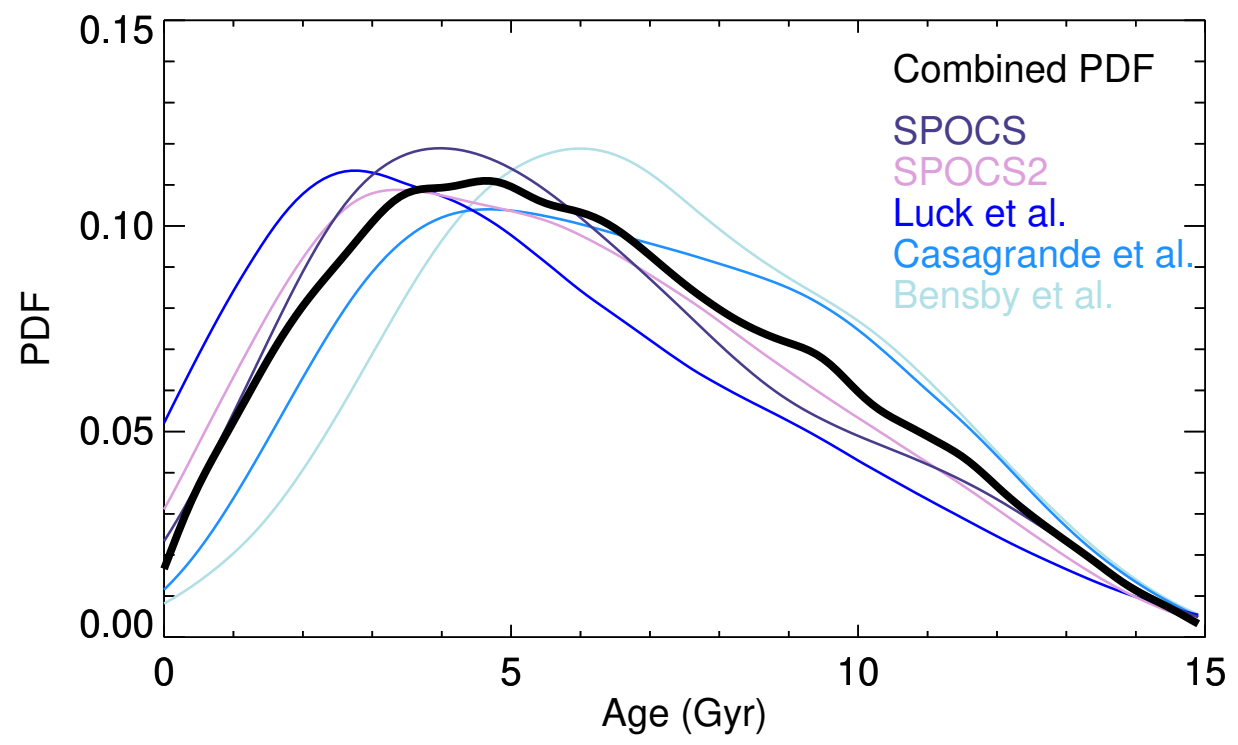

**Figure 2.** The kinematic age probability distribution function (PDF) for L 26-16 A (0003A). The PDF from the individual stellar samples are shown by the colored lines, while the combined PDF is shown by the thick black line. Each PDF is separately normalized so that the area under the curve is one.

The UVW velocity of an object can also be used to ascertain its membership in one of the nearby young moving groups (e.g. Malo et al. 2013; Gagné et al. 2018; Riedel et al. 2017). Once membership has been confidently established, one can then assume the age of the star is the same as the age of the moving group. This provides the strongest constraints on the age of a star, but it is, obviously, only applicable to young stars and nearby groups.

So, for our third approach, we used BANYAN Σ (Gagné et al. 2018) to assess the membership of our M dwarf in nearby young moving groups, using the aforementioned *Gaia* astrometry and, if available, literature radial velocity measurements. Three primaries have a probability >95% of belonging to one of the moving groups considered by BANYAN Σ. 0749A has a 99% probability of belonging to the Ursa Major corona (Gagné et al. 2020); 0541A has a probability of 95% of belonging to the AB Doradus moving group (assuming the photometric distance to the T dwarf as the system's distance); 0558A has a probability of 99% of belonging to the Octans-Near moving group (Zuckerman et al. 2013). We will discuss these three systems in further details in Section 8.

### 7.5. *Color-magnitude diagrams*

Color-magnitude diagrams can be a powerful tool to constrain the age of low-mass stars and brown dwarfs. Since stellar objects shrink during their pre-main se-quence evolution, they move on the color-magnitude diagram from higher luminosity and redder colors to progressively lower luminosities and bluer colors as they age. Figure 3 shows a color-magnitude diagram constructed using the *Gaia* DR3 $G$ and $G_{\rm RP}$ magnitudes, parallaxes, and proper motions. The primaries presented here are compared to the 100 pc field population from the Gaia Catalogue of Nearby Stars (GCNS, Gaia Collaboration et al. 2021), as well as to the empirical sequences derived by Gagné et al. (2021a). Each empirical sequence is constructed combining known members of young moving groups and open clusters of similar age (we refer the reader to Gagné et al. 2021a, for details on how the sequences are constructed). One object stands out from the rest of the sample – 0541A. Its position coincides almost exactly with the Sco-Cen empirical sequence, implying an age of 10–15 Myr for 0541A. This is consistent with its fast rotation (see Section 7.1) but at odds with the lack of H$\alpha$ emission in its spectrum (see Section 7.2). We will discuss the age of this system further in Section 8.6.

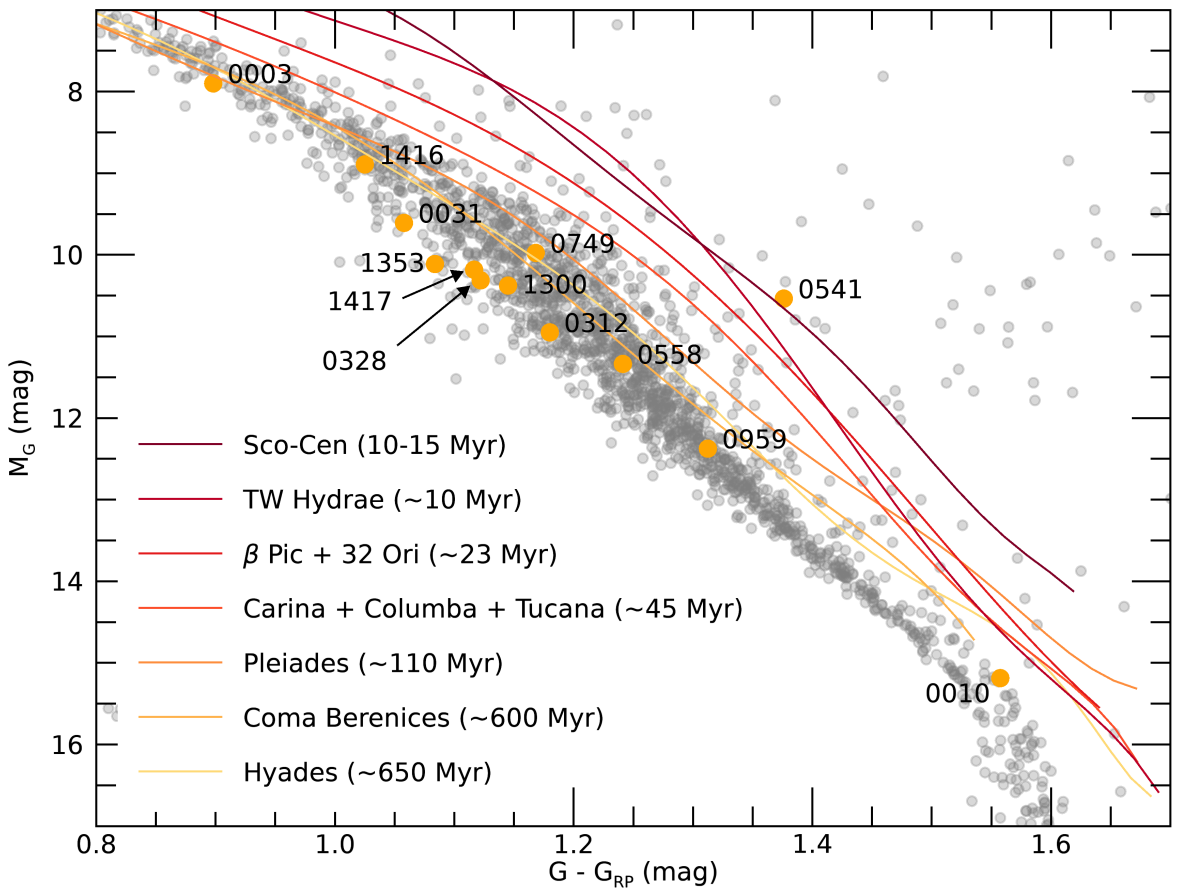

**Figure 3.** Color-magnitude diagram comparing the primaries presented here (orange dots) with the 100 pc field population (grey dots) taken from Gaia Collaboration et al. (2021), and the empirical sequences for various young moving groups and open clusters constructed by Gagné et al. (2021a). To improve the readability of the figure, we dropped the "A" suffix from the M dwarfs labels.

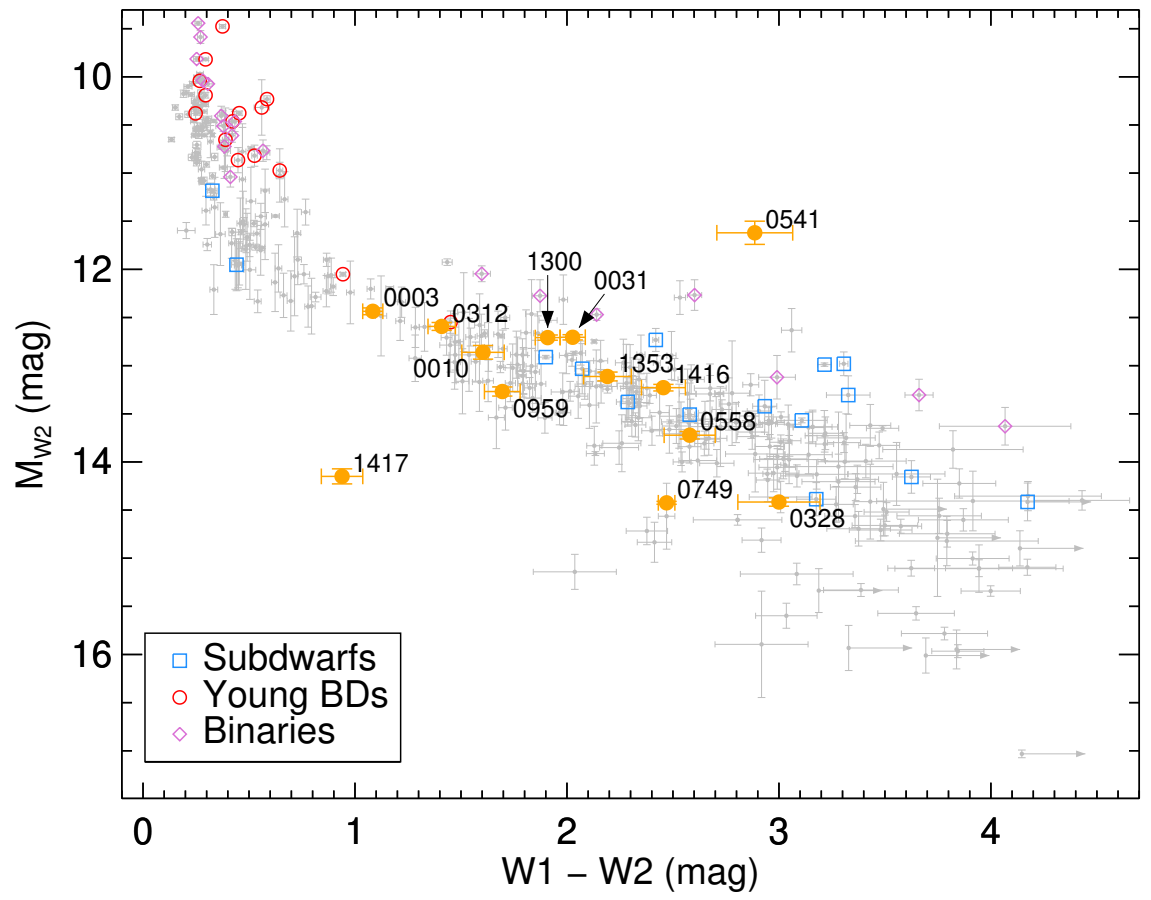

**Figure 4.** Color-magnitude diagram comparing the T dwarfs in our sample (orange dots) with the 20 pc ultracool dwarf census (grey dots) taken from Kirkpatrick et al. (2021a). To improve the readability of the figure, we dropped the "B" suffix from the T dwarfs labels.

Color-magnitude diagrams can also offer hints on the ages of brown dwarfs. Several studies have shown that young L dwarfs are more luminous and redder at near-infrared wavelengths compared to disk-age counterparts of similar mass (e.g. Faherty et al. 2016), while old, metal-poor brown dwarfs are bluer and underluminous (e.g. Zhang et al. 2019).

Figure 4 shows the *WISE* color-magnitude diagram for the 20 pc sample of ultracool dwarfs from Kirkpatrick et al. (2021a) compared to our T dwarfs. To compute the absolute magnitudes for our targets, we assumed that they are at the same distance as their primaries, except for 0749B for which we used its measured parallax. Most T dwarfs in our sample fall close to the locus of disk-age, solar-metallicity objects, but there are a few outliers. 0541B is overluminous by nearly 2 magnitudes compared to objects of similar W1–W2 color, an overluminosity that cannot be explained by unresolved binarity alone. Its primary is very overluminous too (see Figure 3) suggesting that the system could be as young as 10 Myr. 0749B is slightly underluminous, and lies close to a small cluster of objects that are, however, not known to display spectral peculiarities. 0959B and 0328B are also slightly underluminous; however, their position is still consistent with disk-age, solar-metallicity brown dwarfs given the large intrinsic scatter of the T dwarf population. Finally, 1417B is clearly underluminous and bluer compared to solar-metallicity, disk-age T dwarfs. Its position in the color-magnitude diagram is consistent with the old age of its primary, which we found in Section 7.4 to be a likely member of the thick disk. Further details on the individual systems are discussed in Section 8.

## 8. NOTES ON INDIVIDUAL SYSTEMS

### 8.1. *L 26-16 AB (0003AB)*

The primary is photometrically classified in this work as M0 V, making it the earliest type primary in our sample. It was observed by the Radial Velocity Experiment (RAVE DR5, Kunder et al. 2017) which derived $T_{\rm eff} = 4193 \pm 79\,{\rm K}$ (in good agreement with the photometric typing), log $g = 4.82 \pm 0.16$[11], [Fe/H]$= -0.30 \pm 0.09$, RV$= -38.3 \pm 1.1\,\mathrm{km\,s^{-1}}$, and $\mathrm{v\sin i} = 42\,\mathrm{km\,s^{-1}}$. The RAVE spectrum shows Ca H & K in emission, with EW = 0.08 Å, which led the RAVE team to assign the object an age of ∼1.8 Gyr. The *TESS* light curve for the primary does not reveal any obvious variability in either the 30-min cadence data nor the 2-min cadence data. The kinematics provide only weak constraints on the age of the system (3.4–10.6 Gyr). 0003A is detected in *GALEX* NUV with a measured magnitude of 20.44±0.11 mag[12]. No FUV data are available for this region of the sky. With an $R - J$ color of ∼2 mag and $f_{\mathrm{NUV}}/f_{\mathrm{J}} \sim 6.6 \times 10^{-5}$, this object is marginally consistent with the population of young M dwarfs (< 300 Myr) shown in Figure 3 from Shkolnik et al. (2011). Similarly, with NUV−$W1$ ∼12.26 mag and $J - W2 = 0.92$ mag, 0003A is very close to the edge of the "selection box" for young M dwarfs defined by Rodriguez et al. (2013)[13]. The object is undetected by *ROSAT*. Given the lack of strong constraints, we conclude that the age of this system is between 1.8 and 10.6 Gyr.

The newly discovered companion is classified as T4 using our Gemini-S Flamingos-2 spectrum, plotted in Figure 5. The spectrum of the companion does not show any peculiarity (except for a spurious drop in flux at wavelengths $\lesssim 1.1\mu$m due to poor flux calibration), consistent with the age estimate from the primary. We used the Kirkpatrick et al. (2021a) spectral type-to-$T_{\mathrm{eff}}$ relation to estimate the temperature of 0003B, and we then used this estimate with the age constraint for the primary and the isochrones of Baraffe et al. (2003) to estimate its mass. With $T_{\mathrm{eff}}$ ∼1300 K and age in the 1.8–10.6 Gyr range, we obtain a mass in the 52–73 $M_{\mathrm{Jup}}$ range.

### 8.2. *2MASS J00103250+1715490 AB (0010AB)*

The primary is an M8, first discovered by Gizis et al. (2000). Our Keck/NIRES spectrum is shown in Figure 6, and confirms the M8 classification given in the discovery paper. The M dwarf does not fall in any of the *TESS* sectors. The kinematics provide only a weak constraint on the age of this object (2.8–10.1 Gyr). 0010A is undetected in *GALEX*, implying NUV > 22.6 mag and FUV > 22.0 mag. The resulting NUV–W1 color limit (and the J–W2 color) is consistent with the young M dwarf selection criteria of Rodriguez et al. (2013). The object is undetected by *ROSAT*. Given the weak constraint on the age of this M dwarf, we assign an age in the range 2.8–10.1 Gyr.

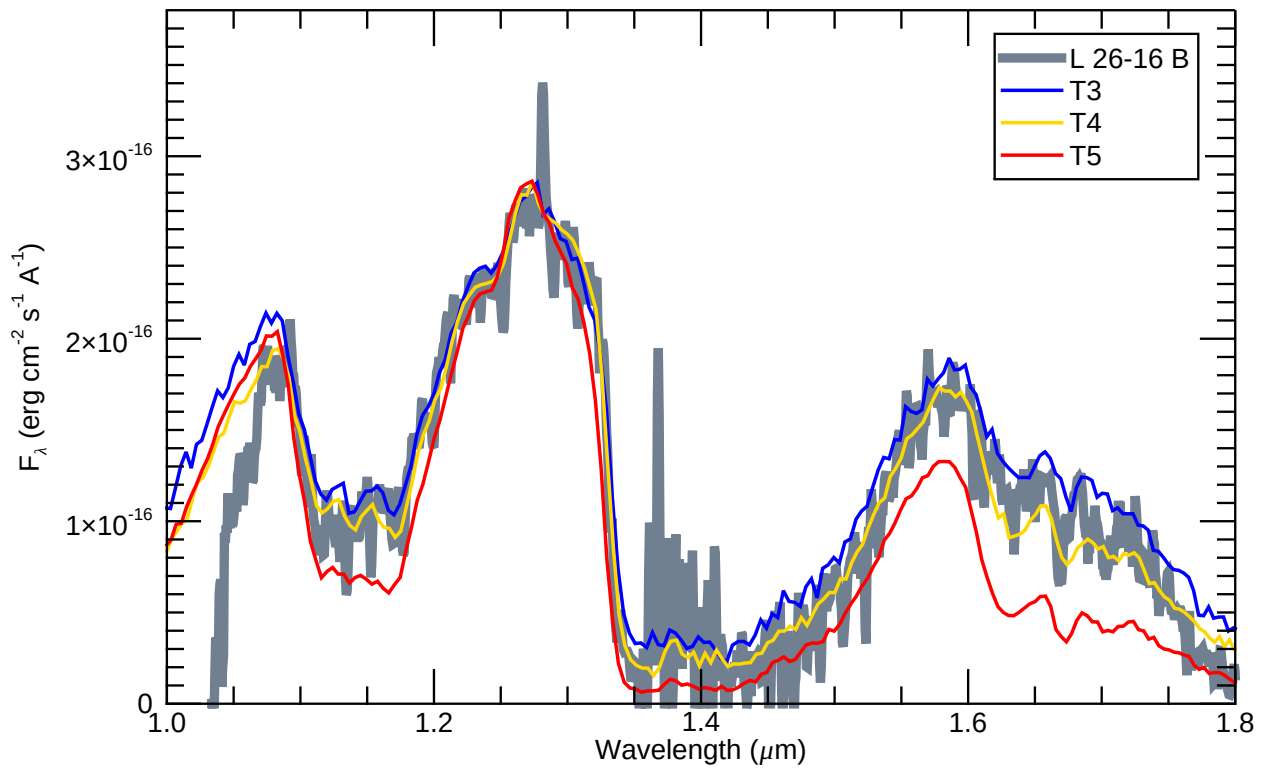

**Figure 5.** The spectrum of L 26-16 B (0003B) obtained with Gemini-S/Flamingos-2, plotted in slate grey. In blue, gold, and red we overplot the T3, T4, and T5 templates. The target spectrum is smoothed with boxcar smoothing of width 3 pixels. The templates are scaled to match the flux of the target at 1.28$\mu$m. Poor flux calibration results in a spurious drop in flux at wavelengths $\lesssim 1.1\mu$m.

The companion spectrum falls between the T5 and T6 spectral templates, with the T6 template fitting the Y- and J-band portion better, and the T5 template providing a better fit at longer wavelength. We therefore classify this object as T5.5. Neither the M dwarf nor the T dwarf show clear peculiarities. Using UHS DR2 data, Schneider et al. (2023) computed a proper motion for this T dwarf of $\mu_\alpha \cos\delta = -182.9 \pm 12.8$ mas yr$^{-1}$ and $\mu_\delta = -39.2 \pm 12.7$ mas yr$^{-1}$, in excellent agreement with the proper motion for the primary. Using the same method described above, we estimate a mass for the companion in the range 50–65 $M_{\mathrm{Jup}}$.

The primary was imaged in a search for wide M7–L8 companions by Allen et al. (2007). They observed 0010A down to a depth of J∼20.5 mag and K∼18.5 mag (MKO) and at separations of 4″ to 32″. They identified the T5.5 as a candidate companion, but ultimately rejected it based on their optical follow-up, which returned an I–J color of 2.2 mag, too blue for objects in the M7–L8 spectral type range of their interest. It is unclear how they obtained such a blue color, since 0010B is undetected in the PS1 3$\pi$ survey, implying i–J$\gtrsim$5.6 mag (which is typical for mid T dwarfs). We speculate that Allen et al. (2007) misidentified the target in their optical image.

### 8.3. *UCAC3 52-1038 AB (0031AB)*

With a projected separation of ∼7100 au, this is, to our knowledge, the fourth widest system containing a T dwarf currently known (Chinchilla et al. 2020).

[11] Throughout this manuscript, $g$ is always expressed in cm s$^{-2}$.

[12] *GALEX* magnitudes are in the AB system.

[13] Rodriguez et al. (2013) select as candidate young K and M stars those with $J - W2 \geq 0.8$ mag and 9.5 ≤NUV–W1< 12.5 mag

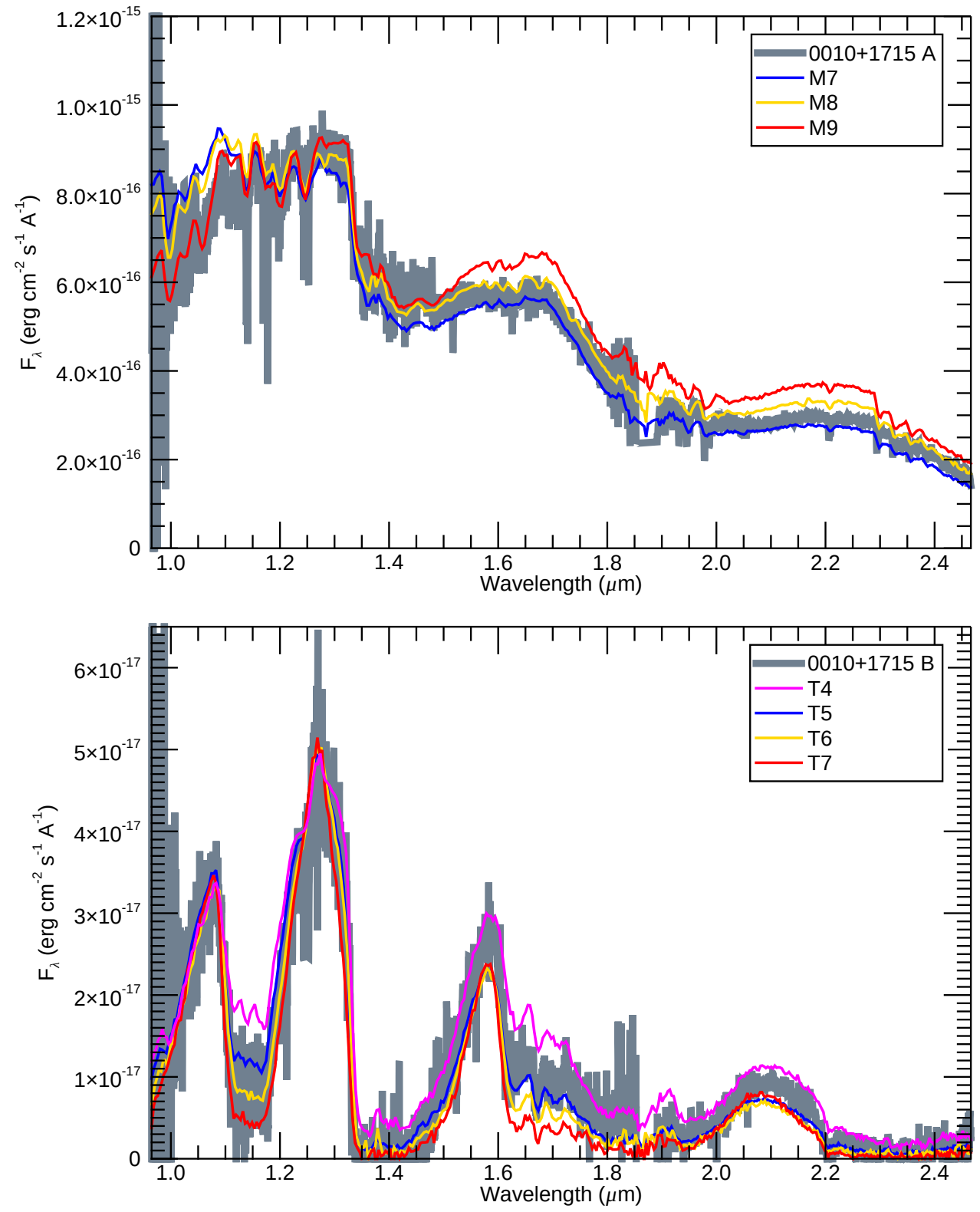

**Figure 6.** The Keck/NIRES spectra of 2MASS J00103250+1715490 A (0010A; top panel) and B (0010B; bottom panel), plotted in slate grey. Spectroscopic standards are plotted in blue, gold, red, and magenta. The targets spectra are smoothed with boxcar smoothing of width 5 pixels.

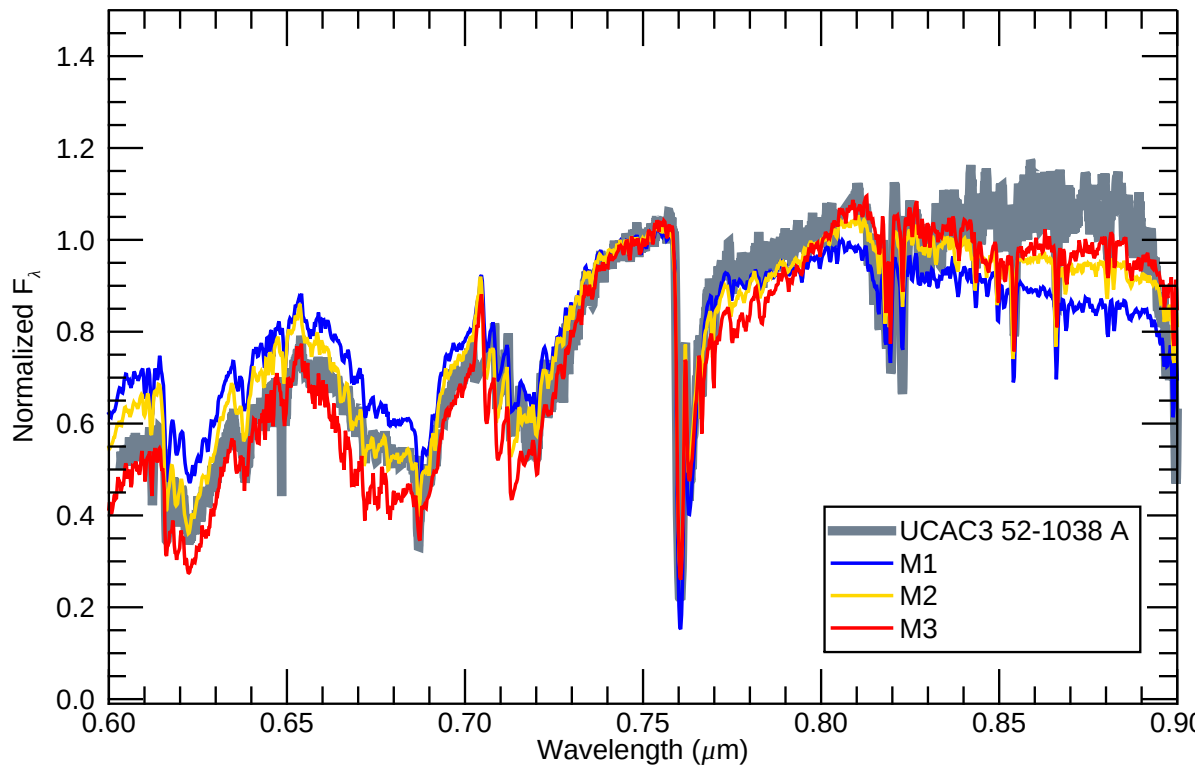

**Figure 7.** The SALT/RSS spectrum of UCAC3 52–1038 A (0031A), plotted in slate grey. We plot spectroscopic standards in blue, gold and red.

The primary is an M2 V based on comparison between the SALT/RSS spectrum and M dwarf spectral templates. As shown in Figure 7, the M2 V template provides an overall good fit to the target spectrum, with the exception of a slight flux suppression of the observed spectrum at wavelengths shorter than $\sim 0.61\,\mu$m and again between $\sim 0.63\,\mu$m and $\sim 0.67\,\mu$m. The M dwarf has been observed by RAVE, and its DR5 entry lists $T_{\rm eff} = 3200$ K, [Fe/H] = -1.25 and an unusual $\log g = 0$, which calls into question the reliability of the measurements. The RAVE RV is 620 km s$^{-1}$, while the $v \sin i$ is 325 km s$^{-1}$ further calling into question the reliability of the measurements. *Gaia* DR2 also provides a measurement of the RV for this object, which is only 15 km s$^{-1}$. Comparison between the SALT spectrum and the sdM2 standard (LSPM J0716+2342; Lépine et al. 2007) rules out a metal poor nature for 0031A. Using the *Gaia* astrometry and RV, we estimate a kinematic age between 2.5 and 9.9 Gyr. The SALT spectrum shows no $H\alpha$, in agreement with the kinematic age estimate. The *TESS* 2-min and 30-min data show no variability. This M dwarf is undetected in the *GALEX* NUV and FUV images, implying limits of 21.6 mag and 21.3 mag, respectively. The resulting NUV–W1 color limit puts this object just outside of the young M dwarf selection box defined by Rodriguez et al. (2013). 0031A is undetected by *ROSAT*. We assign this M dwarf an age in the range 2.5–9.9 Gyr.

The companion is likely a T6 based on its CatWISE2020 Catalog and *Spitzer* photometry. Using the same method described above, we estimate a mass for the companion in the range 43–64 $M_{\rm Jup}$.

### 8.4. *LP 712-16 AB (0312AB)*

The association between the components of this system is uncertain. Visual inspection of the *WISE* images shows the two sources as clearly comoving, and the photometric distance estimate for the T dwarf agrees well with the measured distance for the M dwarf. However, measurements of the proper motion for the T dwarf are highly discrepant with the M dwarf motion, leading to an association probability of 0% if we use the CatWISE2020 measurement, and 0.2% if we use our own fit to the T dwarf coordinates. The proximity between the faint T dwarf and the bright primary (as well as another bright background source) can lead to contamination of its astrometry, so one could argue that neither of the proper motion measurements for the T dwarf are reliable and, therefore, the probability computed with CoMover should be disregarded. While the $\mu_\alpha$ measurements from CatWISE2020 and our own fit are discrepant ($610 \pm 39$ mas yr$^{-1}$ vs. $252 \pm 54$ mas yr$^{-1}$), the $\mu_\delta$ measurements are in good agreement with each other ($-153 \pm 39$ mas yr$^{-1}$ vs. $-150 \pm 51$ mas yr$^{-1}$), as one might expect given that the two bright sources (the M dwarf and the background star) lie to the east and west of the T dwarf, so the right ascension component of the proper motion is the most likely to be corrupted. Moreover, the *Gaia ruwe* for the primary is slightly high (∼1.2), indicating that the astrometry for the primary may be unreliable as well, further complicating the interpretation of this system. If we assume that $\mu_\alpha$ for the T dwarf is the same as for the M dwarf, and use that value along with the measured $\mu_\delta$ as input for CoMover, we obtain a companionship probability of 81%. Dedicated astrometric observations of the T dwarf are advisable to obtain a more reliable proper motion measurement and definitively assess the nature of this pair.

The primary is classified as M4 V using our Lick/Kast spectrum and the M dwarf optical standards defined in Kirkpatrick et al. (2016, and references therein), as shown in Figure 8 (top panel). The data have been telluric corrected, while the standards have not, which explains the discrepancy in the 0.76–0.80$\mu$m region (the telluric Fraunhofer "A" band). We measure the H$\alpha$ equivalent width to be 0.111±0.050 Å which implies $L_{\mathrm{H}\alpha}/L_{\mathrm{bol}} = (3.99 \pm 1.89) \times 10^{-6}$. Both values are well below the activity threshold defined in Kiman et al. (2021). The *TESS* 30-minute and 2-minute light curves appear flat. The kinematic age constraint is 3.3–10.6 Gyr. 0312A is undetected in the *GALEX* NUV and FUV images, implying limits of 21.7 mag and 21.2 mag, respectively. The resulting NUV–W1 color limit is consistent with the young M dwarf selection criteria of Rodriguez et al. (2013). This object is undetected by *ROSAT*. Given this analysis, we assume an age for this M dwarf in the 3.3 to 10.6 Gyr range.

We classify the companion as T6 via comparison of the Keck/NIRES spectrum with templates defined in Burgasser et al. (2006), Kirkpatrick et al. (2010) and Cushing et al. (2011), shown in Figure 8 (bottom panel). While both the T5 and T6 templates reproduce the overall shape of the spectrum, the T6 fits the width of the J-band peak better. Using the same method described above, we estimate a mass for the companion in the range 47–65 $M_{\mathrm{Jup}}$.

### 8.5. *UCAC3 40-6918 AB (0328AB)*

The primary is classified M3 V via comparison between the SALT/RSS spectrum and M dwarf optical templates, as shown in Figure 9. The template matches the SALT spectrum well, with the exception of the same flux suppression seen in the spectrum of 0031A (see Section 8.3). *TESS* 30-minute and 2-minute data are available over multiple sectors, and the overall light curve appears flat. It was observed by RAVE and its DR5 entry reports $T_{\mathrm{eff}} = 3933$ K, $\log g = 4.81$ dex and a somewhat low metallicity [Fe/H] = –0.59 dex. RAVE also reports a radial velocity of 60±1 km s$^{-1}$ and a surprisingly high $v \sin i = 48$ km s$^{-1}$. Žerjal et al. (2017) measured a Ca II infrared triplet equivalent width of 0.27Å, implying a somewhat young age of ∼790 Myr. Our SALT spectrum also shows somewhat stronger Ca II lines with respect to the M3 V template, but the SALT resolution is too low for a definitive assessment. While young M dwarfs rotate faster than their older counterparts, we still find the RAVE $v \sin i$ value to be unlikely. The kinematics, however, suggest an older age (3.3–10.6 Gyr) for this M dwarf, with a 14% probability of belonging to the thick disk. It is undetected in *GALEX* images, and we derive limits to its NUV and FUV magnitudes of 21.9 mag and 21.4 mag, respectively. The NUV–W1 limit places this object outside the young M dwarf locus in Rodriguez et al. (2013). The source is undetected by *ROSAT*. Given the inconsistency between the Žerjal et al. (2017) age estimate and other age indicators considered here, we conservatively assign an age to this object in the 0.79–10.6 Gyr range.

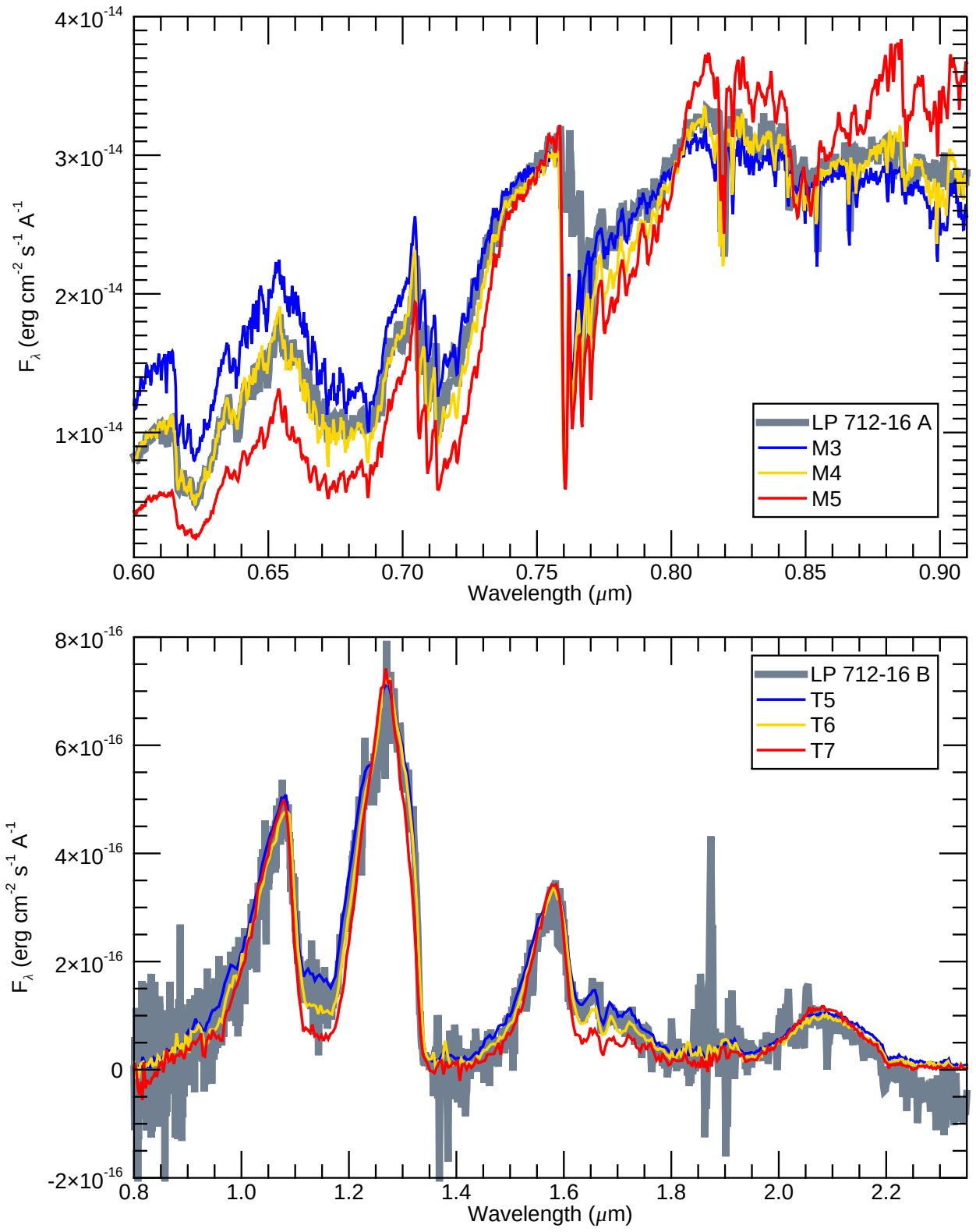

**Figure 8.** The Lick/Kast spectrum of LP 712-16 A (0312A; top panel) and the Magellan/FIRE spectrum of LP 712-16 B (0312B; bottom panel), plotted in slate grey. In each panel, we plot spectroscopic standards in blue, gold and red.

The T dwarf is classified as T8 based on its CatWISE2020 and *Spitzer* photometry. If we assume an age of 790 Myr, as implied by the Ca II equivalent width, the companion would have a mass of only 15$M_{\mathrm{Jup}}$. If instead we assume the upper limit on the age of the M dwarf (10.6 Gyr), the mass of the T dwarf would be 46 $M_{\rm Jup}$.

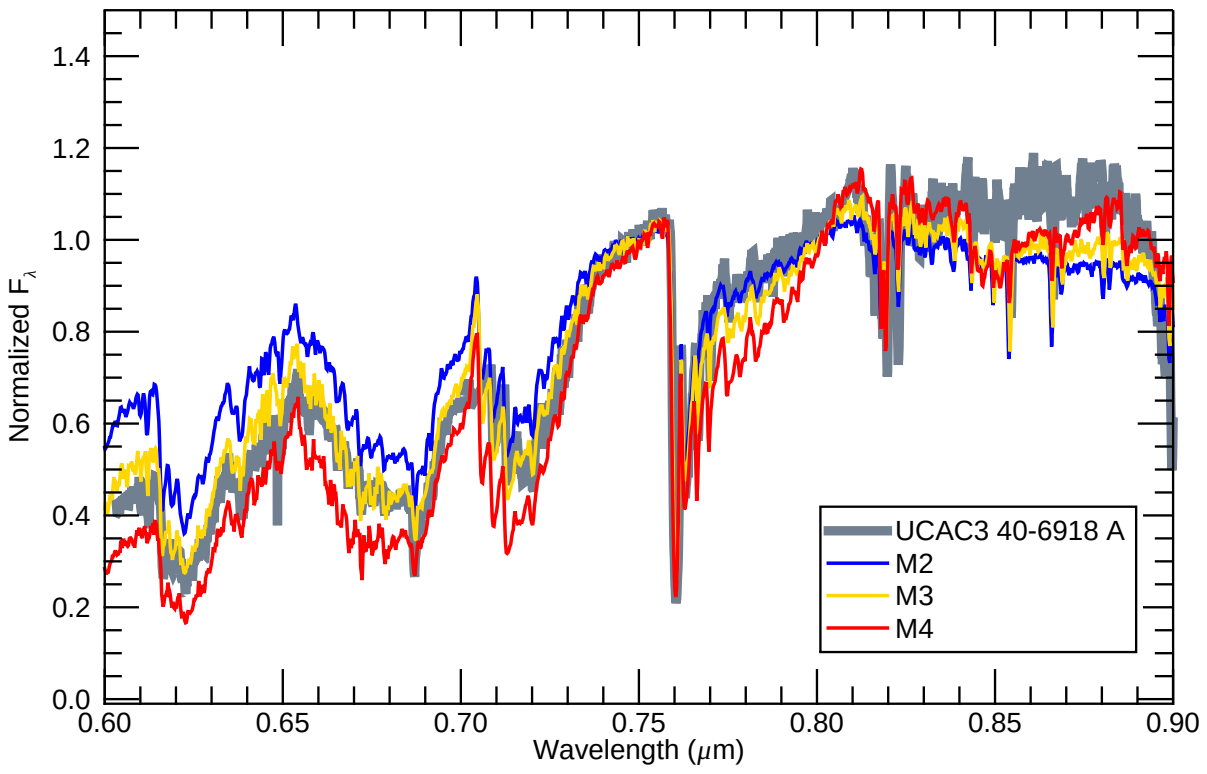

**Figure 9.** The SALT/RSS spectrum of UCAC3 40–6918 A (0328A), plotted in slate grey. We plot spectroscopic standards in blue, gold and red.

### 8.6. *CWISE J054129.32–745021.5 AB (0541AB)*

The association between the primary and its putative companion is questionable. Visual inspection of the WISE images shows the two sources as clearly comoving, and the proper motion measurements of the primary and companion are in good agreement, albeit with large uncertainties on the companion motion measurement (see Table 4). The photometric distance of the T dwarf and the parallax measurement for the M dwarf are, however, highly discrepant, with the T dwarf at a distance of ∼34 pc and the M dwarf at a distance of ∼77 pc. The resulting association probability returned by CoMover is, therefore, 0%. However, we note that the *ruwe* on the M dwarf measurements is 15.192, which calls into question the reliability of the *Gaia* parallax. Additional clues come from the comparison between the *Gaia* DR3 and DR2 astrometry. While the change in measured distance is negligible (0.5 pc, i.e. 0.3$\sigma$), the acceleration in the proper motion is significant, with $\Delta\mu_{\rm tot} = 4.73$ mas yr$^{-1}$ or 4.5$\sigma$. Moreover, the *Gaia* DR2 *ruwe* is smaller than the DR3 value (8.919 vs. 15.192). Proper motion acceleration and increase in *ruwe* are known to be strong evidence of unresolved binarity (Penoyre et al. 2022a,b). Finally, both astrometric distance measurements are discrepant with the photometric distance for the M dwarf (∼57 pc, based on the spectral type and the available photometry). If we disregard the *Gaia* parallax for the M dwarf and use its photometric distance instead, CoMover computes an association probability of 99.4%. Future *Gaia* data releases will hopefully solve the mystery by providing improved astrometry for the primary.

We classify the primary as M4 V based on comparison between the SALT spectrum and M dwarf spectral standards (Figure 10). The template provides a very good match to the full spectrum, with no clear sign of peculiarity. 0541A is the fastest rotator in the sample, with a rotation period measured from the *TESS* 2-min light curve of ∼0.13 d. Such a short rotation period hints at a very young age for this system. Comparison with typical rotation periods of young moving groups, as shown in Figure 1, suggests that 0541A could be younger than the Pleiades (120 Myr; Rebull et al. 2016). However, there is no noticeable $H\alpha$ in the optical spectrum (Figure 10). The object is undetected in both *GALEX* bands, leading to 3$\sigma$ limits of NUV>21.6 mag and FUV>21.0 mag. These non-detections, however, do not rule out the possibility that 0541A is a young M dwarf, since the resulting J–W2, NUV–W1, R–J colors and limit and the $f_{\rm NUV}/f_{\rm J}$ limit still place this object within the loci of young M dwarfs defined in Shkolnik et al. (2011) and Rodriguez et al. (2013). The kinematics do not provide strong constraints on the age of this system, with a broad age range estimate of 2.9–10.3 Gyr, and we remind the reader that our sample only contains stars older than 2 Gyr, so the kinematic analysis would be insensitive to younger ages. On the other hand, the high *ruwe* value and proper motion acceleration reported above could be an indication that 0541A is an unresolved binary, and interaction between the two components could be responsible for the very high rotation rate observed, instead of youth. All things considered, we tentatively assign this object an age <1 Gyr.

The companion is a T8 based on its CatWISE2020 W1 and W2 magnitudes and the calibrations of Kirkpatrick et al. (2019). Near-infrared spectroscopy is desirable to further characterize this object. If we assume the system is younger than 1 Gyr, the T8 would have a mass < 16 $M_{\rm Jup}$, making this likely to be a wide planetary-mass companion. If we assume an age as young as 10 Myr, as implied by the position of the M dwarf in Figure 3, the T dwarf would have a mass of ∼1.8 $M_{\rm Jup}$.

### 8.7. *2MASS J05581644–4501559 AB (0558AB)*

The primary is an M4 V according to our template fit to the SALT spectrum, as shown in Figure 11. The M4 V template provides an excellent fit to the >0.7 $\mu$m portion of the spectrum, while the 0.6–0.7 $\mu$m portion is a poor fit. The spectrum shows clear $H\alpha$ in emission, with an equivalent width of 2.323 Å, indicative of youth. The discrepancies between the M4 V template and the observed spectrum can also be explained by youth. Young M dwarfs tend to have shallower TiO, CaH, and FeH absorption bands (Gray & Corbally 2009, and references therein), as tentatively observed in the spectrum of 0558A (the actual depth of the TiO and CaH bands in the 0.6–0.7 $\mu$m range cannot be measured since the band head at 0.7 $\mu$m falls in the gap between the three detectors). The *TESS* 2-minute and 30-minute light curve shows clear variability with a period of 1.56 d, and several flares. Such a short rotation period and the frequency of flares are consistent with a young age for 0558A. In Figure 1 this object is somewhere between the Pleiades (∼120 Myr) and the Praesepe (∼650 Myr) populations, albeit in a region of large scatter and significant overlap between the two. Our M dwarf does, however, clearly show a much shorter rotation period than the old, field population. We therefore conservatively assume that 0558A has an age ≤650 Myr.

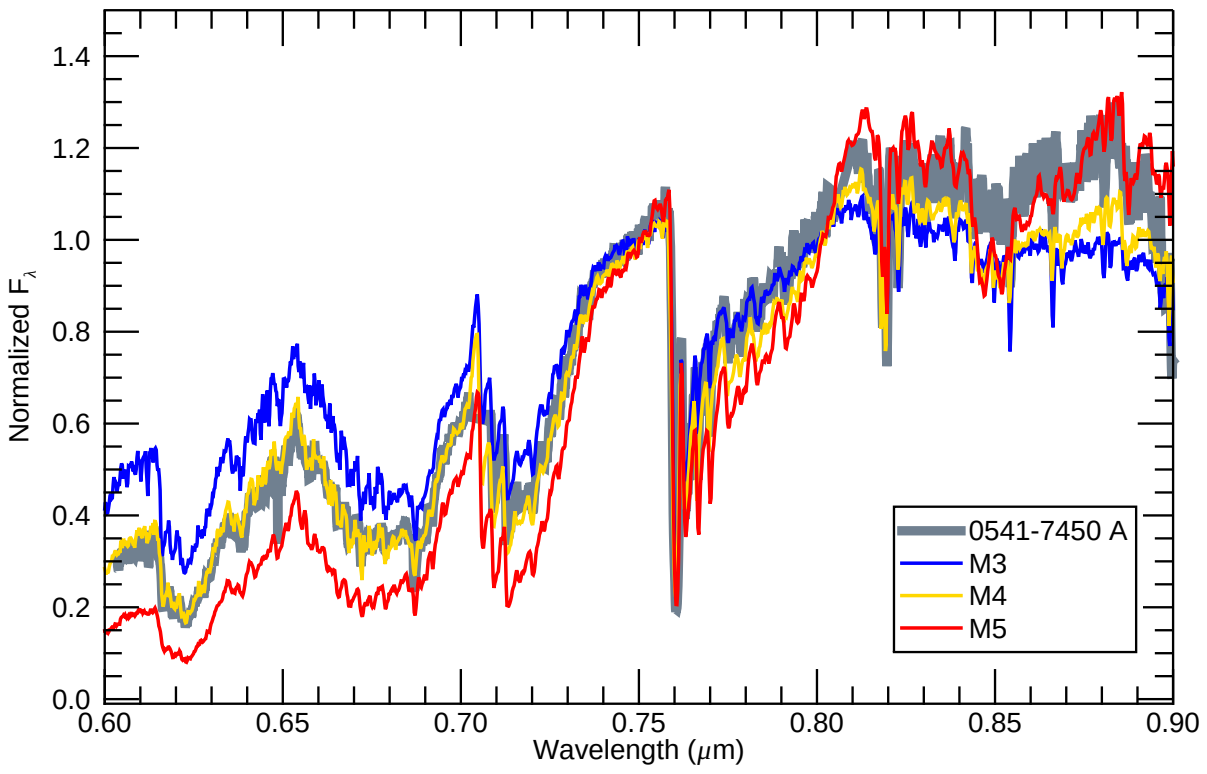

**Figure 10.** The SALT/RSS spectrum of CWISE J054129.32–745021.5 A (0541A), plotted in slate grey. Spectroscopic standards are plotted in blue, gold and red.

The youth of the primary can be investigated further with its UV brightness and X-ray emission. The target is detected in the near-UV band of *GALEX* with an NUV magnitude of 21.46±0.23 mag, while it is undetected in the far-UV band, resulting in an upper limit of > 21.0 mag. For 0558A we use the NUV flux from *GALEX* and the J-band flux from 2MASS to obtain $f_{\rm NUV}/f_J = 9.19 \times 10^{-5}$, and the R magnitude from the USNO-B1.0 catalog (Monet et al. 2003) to obtain R–J = 3.57 mag. Comparing our target with objects in Figure 3 of Shkolnik et al. (2011) we find it sits in an area of large scatter, but overall below objects that those authors considered young M dwarf candidates. We find 0558A has NUV–W1 = 11.976±0.235 mag, and J–W2 = 1.185±0.027 mag, which passes the selection criteria for young M dwarfs defined by Rodriguez et al. (2013). Overall, the UV analysis is consistent with our assumption of an age ≤650 Myr.

Inspection of the *ROSAT* image of the field reveals that the source sits inside a large "blob" of faint X-ray emission, as shown in Figure 12. The *ROSAT* all-sky survey bright source catalogue lists one source, 1RXS J055818.1–450146, located approximately 20″ away from the *Gaia* coordinates for *0558A*, with a brightness of 2.33 ± 0.72 × $10^{-2}$ counts per second. The count rate is converted to flux using the formula $F_X$ = (count rate) × (8.31 + 5.30 × HR1) × $10^{-12}$ erg cm$^{-2}$ s$^{-1}$, where HR1 = $(B-A)/(B+A)$, with A being the counts in the soft channel (0.1-0.4 keV) and B being the counts in the hard channel (0.5-2.0 keV). This conversion formula was derived by Schmitt et al. (1995) for nearby K and M dwarfs and is, therefore, appropriate for our primaries. Assuming the *ROSAT* detected flux comes from 0558A, the resulting ratio between X-ray and infrared brightness is $\log F_{\rm X}/F_J = -2.25$, which can be compared with young and field objects. Looking at Figure 3 in Shkolnik et al. (2009) we see that 0558A appears clearly brighter in X-rays than old, field objects of similar I–J color (1.32 mag). A value of $\log F_{\rm X}/F_J = -2.25$ is indeed close to the saturation limit for M dwarfs, indicative of youth. Our X-ray analysis is therefore also consistent with an age ≤650 Myr. A nearby, bright UV source (easily identified in Figure 12 to the north-west of the target) and several faint UV sources in the field could however contribute to the X-ray flux, or even be solely responsible for the *ROSAT* detection.

Our comparison to the kinematic sample (Section 7) yields an age in the 2.4–9.6 Gyr, at odds with the above assumption, but since our kinematic sample is composed of stars with age >2 Gyr, our kinematic analysis is insensitive to ages younger than that. Finally, we used BANYAN Σ to assess the membership of 0558A to nearby young moving groups, and found a probability of 99% of belonging to the 30-100 Myr old Octans-Near moving group (Zuckerman et al. 2013). The nature of this proposed moving group, however, is disputed (Mamajek 2016). We conclude that 0558A is likely younger than 650 Myr, but most likely older than the Pleiades, and therefore assign this object an age of 120–650 Myr.

We obtained a Magellan/FIRE spectrum for the companion and classify it as T8 based on template comparison, as shown in Figure 11. The target does not show strong peculiarities, except for a slightly enhanced K-band spectrum, which appears brighter than the T8 standard. While an enhanced K-band (and H-band) flux is typically associated with youth in L dwarfs (see e.g. Faherty et al. 2016), this relation is not well established in T dwarfs, and, in fact, the 0749AB system may completely subvert this expectation (see Section 8.8). Given the age of the primary (120–650 Myr), the companion has a mass of 6–12 $M_{\rm Jup}$, below the minimum mass for deuterium burning.

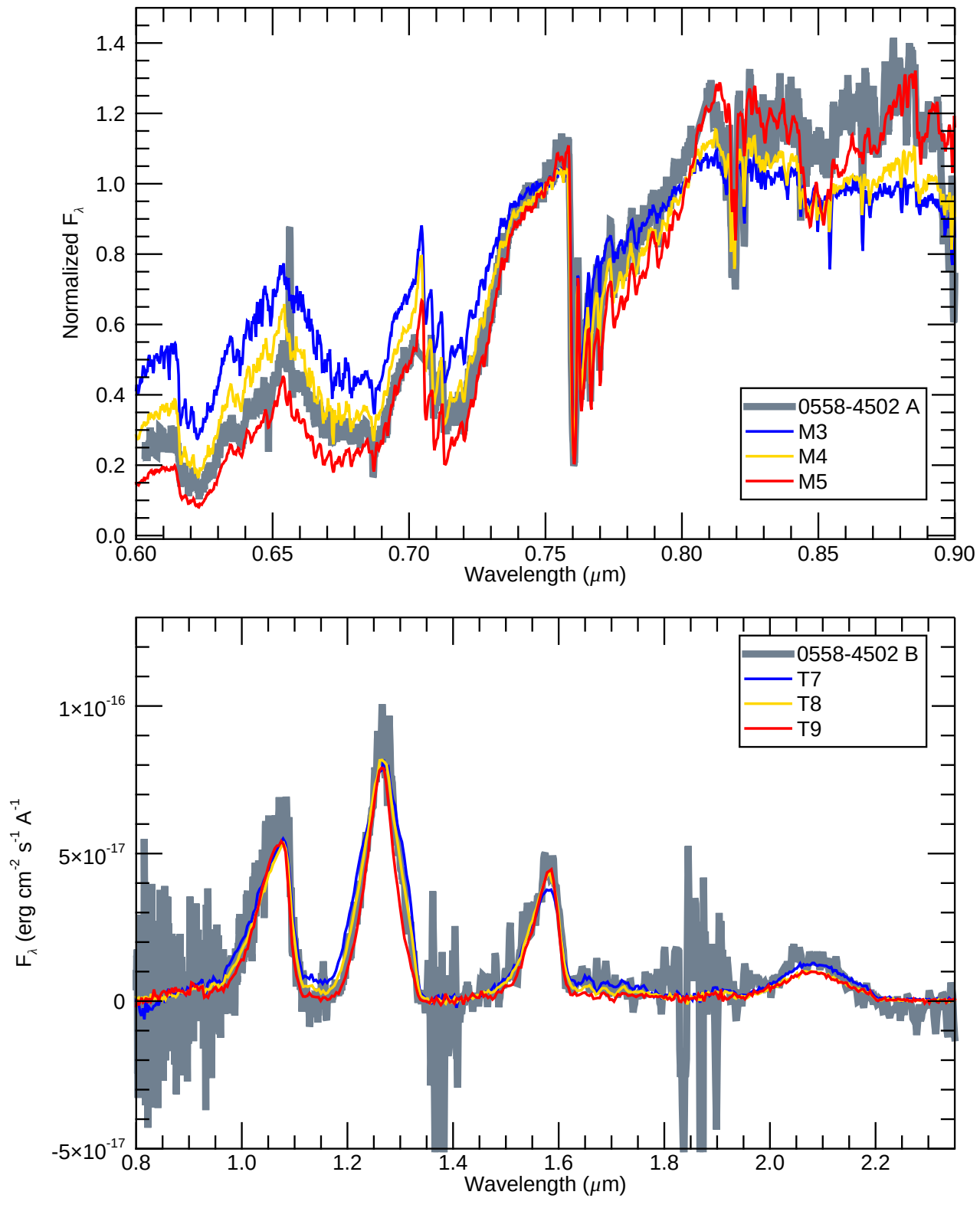

**Figure 11.** The SALT/RSS spectrum of 2MASS J05581644–4501559 A (0558A; top panel) and the Magellan/FIRE spectrum of 2MASS J05581644–4501559 B (0558B; bottom panel), plotted in slate grey. In each panel, spectroscopic standards are plotted in blue, gold and red. The spectrum of 0558B is smoothed with boxcar smoothing of width 3 pixels.

### 8.8. *L 34-26 + WISE 0751–7634 (0749AB)*

The association between these two previously-known objects has been independently reported by Zhang et al. (2021b). With an angular separation of $\sim$597″ and a projected separation of $\sim$6500 au, this is one of the widest systems containing a T dwarf currently known.

The primary is the well known M3 V L 34-26 (Torres et al. 2006). The star is known to be rotationally variable with a period of 2.827 days, which led Newton et al. (2016) to constrain the age of this object to $<$ 2 Gyr. *TESS* data confirms the rotation variability and the known period, and reveals numerous flares, consistent with youth. Our analysis of Figure 1 leads to more stringent age constraints. 0749A is clearly a much faster rotator than the field population, but also appears to rotate faster than the majority of Praesepe M dwarfs. Large scatter, however, prevents us from placing any more stringent constraints on the age of this object based on rotation alone, so we adopt a somewhat conservative upper limit of 650 Myr.

The kinematics of this M dwarf help us constrain its age further. Using BANYAN $\Sigma$ we obtain a 99% probability for this object to belong to the proposed Ursa Major corona Gagné et al. ($\sim$400 Myr; 2020). However, 0749A is not close to the Ursa Major core, and the proposed corona has relatively wide U, V, and W distributions, which makes it susceptible to contamination.

A young age is further substantiated by the position of this object in the CMD of Figure 3, where it falls almost exactly on the 400 Myr empirical sequence.

$H\alpha$ equivalent width measurements in the literature range from $\sim$3 to $\sim$8 Å, which is consistent with an age of $<$1 Gyr.

The object is well detected in *GALEX*, with FUV = 19.797±0.15 mag and NUV = 18.101±0.035 mag, which leads to $f_{\rm NUV}/f_J = 1.156 \times 10^{-4}$, and NUV-W1 = 11.395±0.038 mag. This places the object among the young candidates of Shkolnik et al. (2011) and Rodriguez et al. (2013).

The X-ray photometry from *ROSAT* and *XMM-Newton* yield $\log L_{\rm X}$ (erg s$^{-1}$) = 28.92, $\log L_{\rm X}/L_{\rm bol} = -3.15$, and $\log F_{\rm X}/F_J = -2.12$, a larger ratio than all of the Hyades (650 Myr) M dwarfs shown in Figure 3 of Shkolnik et al. (2009). This is consistent with our assumed age constraint.

This M dwarf was observed with ESPRESSO by Hojjatpanah et al. (2019), who measured $v \sin i = 7.40 \pm 0.42$ km s$^{-1}$. Assuming the typical radius of an M3 V[14], the measured $v \sin i$ and *TESS* period, we find with the publicly available *sini_mcmc* python program[15] that 0749A is seen nearly equator on (i = 81.8±5.8 deg).

The T9 dwarf companion, WISE 0751–7634, is also a well known, well studied object (e.g. Kirkpatrick et al. 2012; Leggett et al. 2015; Kirkpatrick et al. 2019). It is known to be underluminous in $M_H$, $M_{\rm W1}$, and $M_{\rm ch1}$ with respect to objects of similar ch1–ch2 color (e.g. Kirkpatrick et al. 2021a). Another object that shows similar peculiarities is the old, metal-poor sdT8 WISE 2005+5424, which might indicate that 0749B is also a metal-poor sdT. Several other recently discovered cold substellar subdwarfs appear unusually blue (Schneider et al. 2020; Meisner et al. 2021), and sometimes extremely blue (Kirkpatrick et al. 2021b), in J–ch2 and J–W2 with respect to objects of similar ch1–ch2 and W1–W2, further strengthening the hypothesis that 0749B is an old, metal-poor sdT. Its companionship to 0749A, however, calls into question this assumption, and opens

[14] taken from https://www.pas.rochester.edu/~emamajek/EEM_dwarf_UBVIJHK_colors_Teff.txt

[15] https://github.com/jgagneastro/sini_mcmc/blob/master/sini_mcmc.py

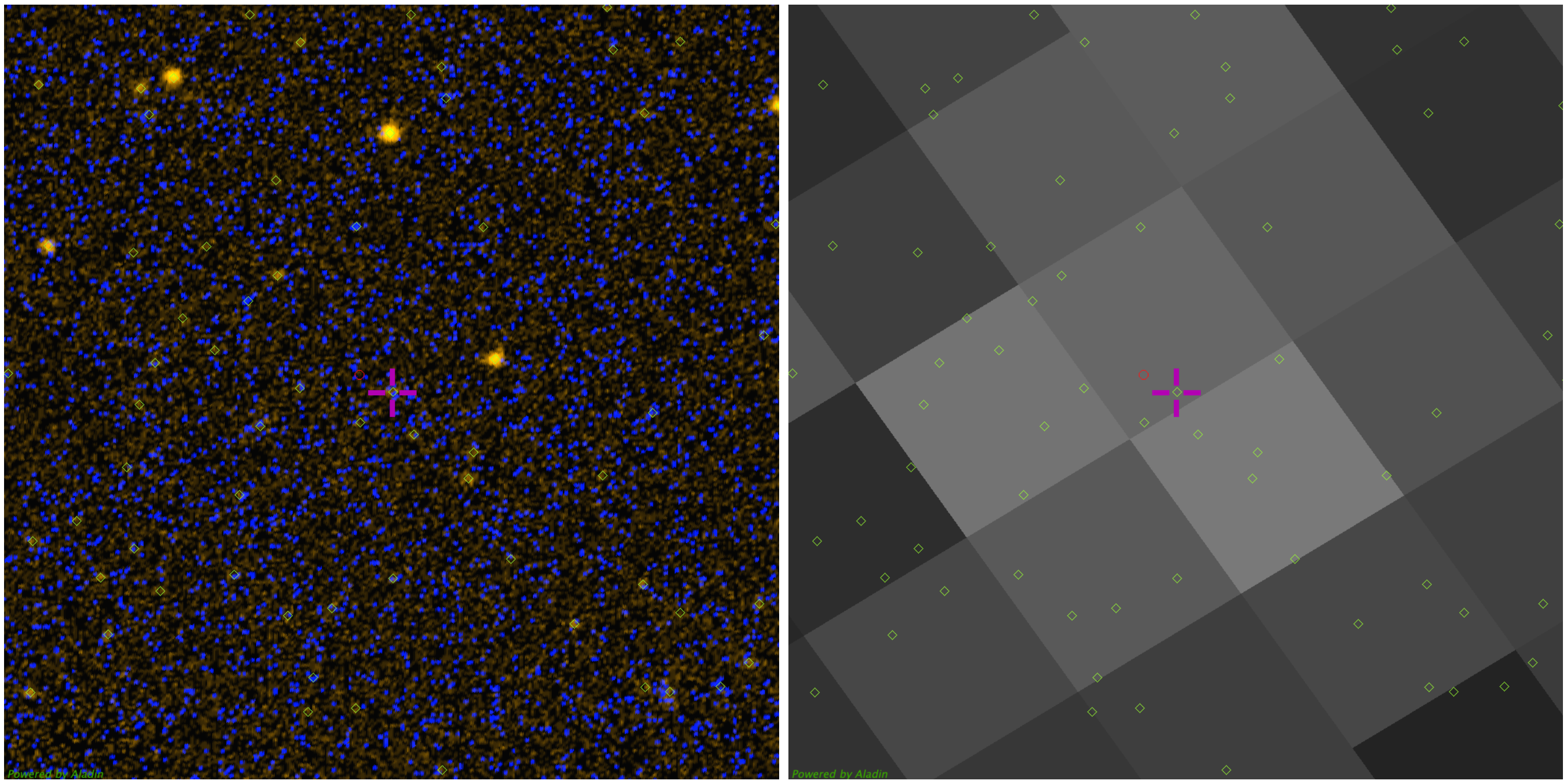

**Figure 12.** *GALEX* false-color image (left) and *ROSAT* (right) image centered around the *Gaia* position for 0558A. In the *GALEX* image, the near-UV channel is plotted in red and the far-UV channel is plotted in blue. In each panel a purple cross marks the *Gaia* coordinates for 0558A, green diamonds mark all UV sources detected by *GALEX*, while the red circle marks the position of the X-ray source 1RXS J055818.1–450146. Both panels are $7' \times 7'$ and oriented with north up and east to the left.

numerous intriguing questions on the nature of this system.

As discussed in the previous paragraphs, 0749A is a young star, and is not metal poor, with the high-resolution spectroscopic survey of Hojjatpanah et al. (2019) measuring [Fe/H] = $0.00 \pm 0.08$ dex. Young L dwarfs with well constrained ages have been shown to be unusually red and overluminous in the near-infrared by numerous studies (e.g. Faherty et al. 2016; Liu et al. 2016, and references therein). The redness of their spectra is most likely caused by the reduced $H_2$ collision-induced absorption (CIA) due to the low gravity in their atmosphere as well as enhanced iron- and silicate-dust content in the upper layers of the photosphere (Marocco et al. 2014; Hiranaka et al. 2016; Burningham et al. 2021). It has been assumed this trend would continue down to lower temperatures, for example in the case of the suspected young T dwarf CFBDSIR 2149-0403 (Delorme et al. 2017). Atmospheric models for low-gravity T dwarfs also imply redder colors and a redder SED for $\log g < 4.0$ objects, with $\log g$ being a proxy for age since brown dwarfs shrink as they age.

What is, therefore, the nature of the 0749AB system? The companionship probability given by CoMover is 100%, so we can confidently rule out the possibility that the two objects are simply an unfortunate case of chance alignment.

A possibility is that 0749B formed as a planet within the protoplanetary disk of 0749A. The peculiar, puzzling observed properties of the T9 could then be explained as arising from a non-solar C/O ratio. The C/O ratio of a planet is known not to reflect the host star C/O ratio (which is also the "average" protoplanetary disk C/O ratio), as it is influenced by localized enhancement/depletion of carbon- and oxygen-bearing molecules within the disk (e.g. Öberg et al. 2011; Najita et al. 2013). However, the system has a projected separation of nearly 6,500 au, so formation in-situ is impossible. 0749B would have had to form in the inner part of the disk and subsequently migrate outward for thousands of au. Numerical simulations by Veras et al. (2009) have shown that planet-planet interactions can lead to planetary systems with massive planets in very wide orbits (up to $10^5$ au) which are stable enough to remain bound for hundreds of millions of years. However, if we assume an age of $\sim$650 Myr, 0749B has a mass of $\sim 8\,M_{\rm Jup}$. While planet-like formation is not impossible, the formation of more than one massive planet (the Veras et al. 2009 simulations assume at least three super-Jupiters in the system) around an M3 V is highly unlikely, given that the protoplanetary disk would not have had enough material. Another possible cause of migration is a "stellar fly-by", i.e. a close encounter between the 0749AB system and another star. However, in the case of a planetary system–single star interaction, numerical simulations show that, if the planetary system survives the encounter, the semi-major axis only increases by approximately 25% (depending on the relative velocity of the encounter; Fregeau et al. 2006). More complex scenarios, such as binary-binary or planetary system-planetary system encounters, could lead to different outcomes (Yip et al. 2023), including exchange of components between the original binaries, or the formation of stable hierarchical triple systems (e.g. Malmberg et al. 2011). Overall, stellar fly-bys could be responsible for the outward migration of 0749B. Atmospheric retrieval and determination of the C/O ratio for both components of this system is therefore desirable to shed light on this hypothesis.

An alternative scenario would be that the M dwarf is not actually young, with its apparent youth being the result of tidal and/or magnetic interactions with an unseen companion. Star-planet interactions are known to cause a spin-up of the host star and an increase in its chromospheric activity (Vidotto 2020, and references therein). In this scenario, the 0749AB system would be old, and both components formed through the stellar binary formation pathway. The M3 V would appear to rotate faster and to be unusually active because of its interaction with a planet and/or brown dwarf in a tight orbit around it. The *TESS* light curve rules out the presence of a transiting companion, but *Gaia* reports a "spectroscopic binary" probability of 41% (*classprob_dsc_specmod_binarystar*), and a significant astrometric excess noise of 0.168 mas (31.8$\sigma$), which favors this interpretation. Further RV monitoring and/or high contrast imaging are desirable to identify a possible companion. The possible membership of the primary to the proposed Ursa Major corona, however, is at odds with this interpretation.

Finally, we speculate that this system could be the result of a capture. 0749A would indeed be a young M3 V, 0749B would indeed be an old, low-metallicity T9, and the association between the two would be the result of an aforementioned fly-by resulting in the capture of 0749B by 0749A. Stellar fly-bys are known to be common even in the Galactic disk, with Bailer-Jones et al. (2018) estimating 19.7±2.2 Myr$^{-1}$ encounters within 1 pc for the Solar system. For a fly-by to result in a capture, however, the relative velocity of the encounter must be very low (Fregeau et al. 2006), a condition that may be hard to meet in an encounter between a young, kinematically "cold" star like 0749A and the older, kinematically "heated" 0749B.

### 8.9. *SCR J0959-3007 AB (0959AB)*

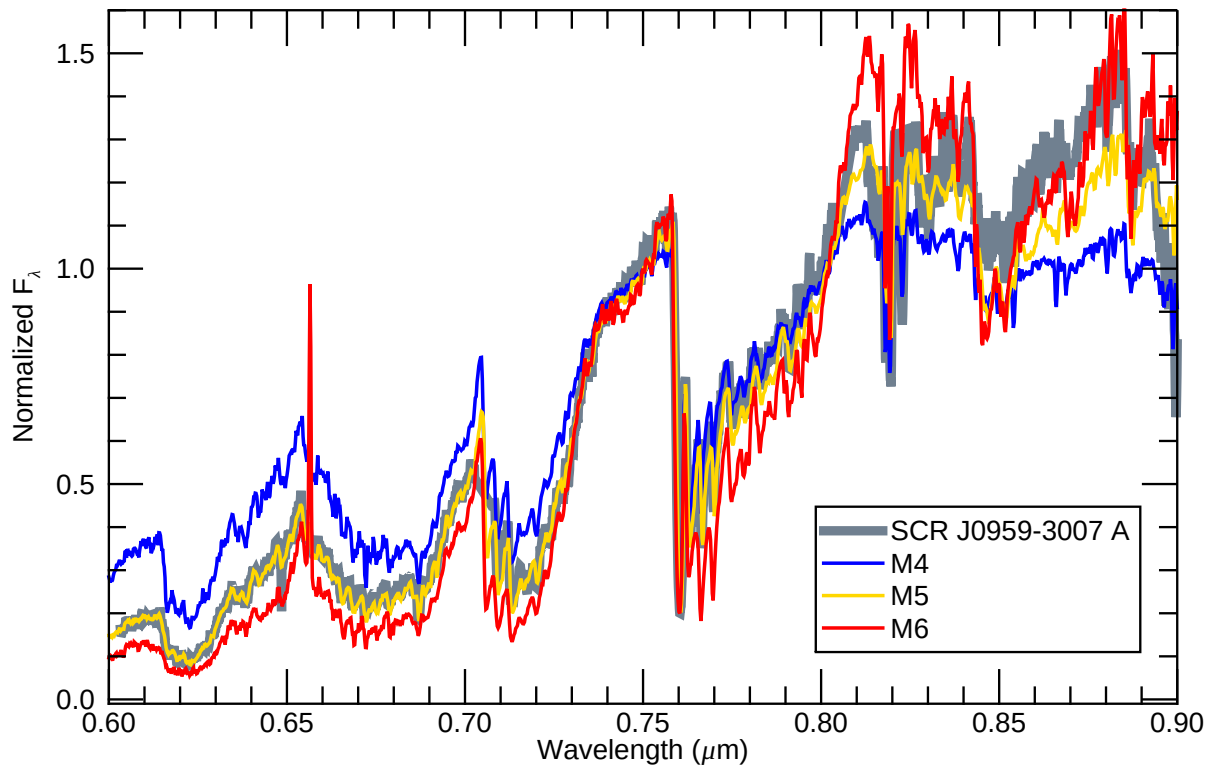

**Figure 13.** The SALT/RSS spectrum of SCR J0959-3007 A (0959A), plotted in slate grey. The M4, M5, and M6 spectroscopic standards are plotted in blue, gold and red.

We classify the primary as M5 V based on comparison between our SALT spectrum and M dwarf templates, as shown in Figure 13. The template matches very well the observed spectrum, with no sign of peculiarity. The 30-minute data from *TESS* shows variability with a period of 3.1 days. In the color-period plot of Figure 1, 0959A is just above the Praesepe sample, albeit in a part of the plot that shows very large scatter. The M dwarf is, however, clearly a faster rotator than the field M dwarfs, suggesting it is younger than ~1 Gyr. The target is detected in the NUV band of *GALEX* with NUV = 22.40±0.47 mag, while it is undetected in the FUV band, implying FUV>21.4 mag. The resulting $f_{\rm NUV}/f_J$ is low compared to objects of similar R–J color in Figure 3 of Shkolnik et al. (2011), and the NUV-W1 and J–W2 colors place it well within the selection criteria for young objects in Rodriguez et al. (2013). These results are consistent with a somewhat young age of <1 Gyr. Kinematic information does not help us constrain the age of the system any further, so we adopt <1 Gyr as our best estimate.

We classify the companion as T5 based on its CatWISE2020 photometry. Assuming an upper limit on the age of 1 Gyr, the corresponding upper limit on the mass of the companion is 35 $M_{\rm Jup}$.

### 8.10. *UCAC4 307-069397 AB (1300AB)*

The primary is an M4 V based on template fitting of the Kast spectrum, as shown in Figure 14. The *TESS* light curve (30- and 2-minute) appears featureless. Using the Kast spectrum we measure the equivalent width of $H\alpha$ to be 0.186±0.050 Å, well below the activity threshold of 0.82Å defined by Kiman et al. (2021). The primary is undetected in *GALEX*, leading to FUV>21.1 mag and NUV>22.1 mag. The only age constraint we have on this system comes from our kinematics analysis, which yields an age estimate in the 3.3–10.5 Gyr range.

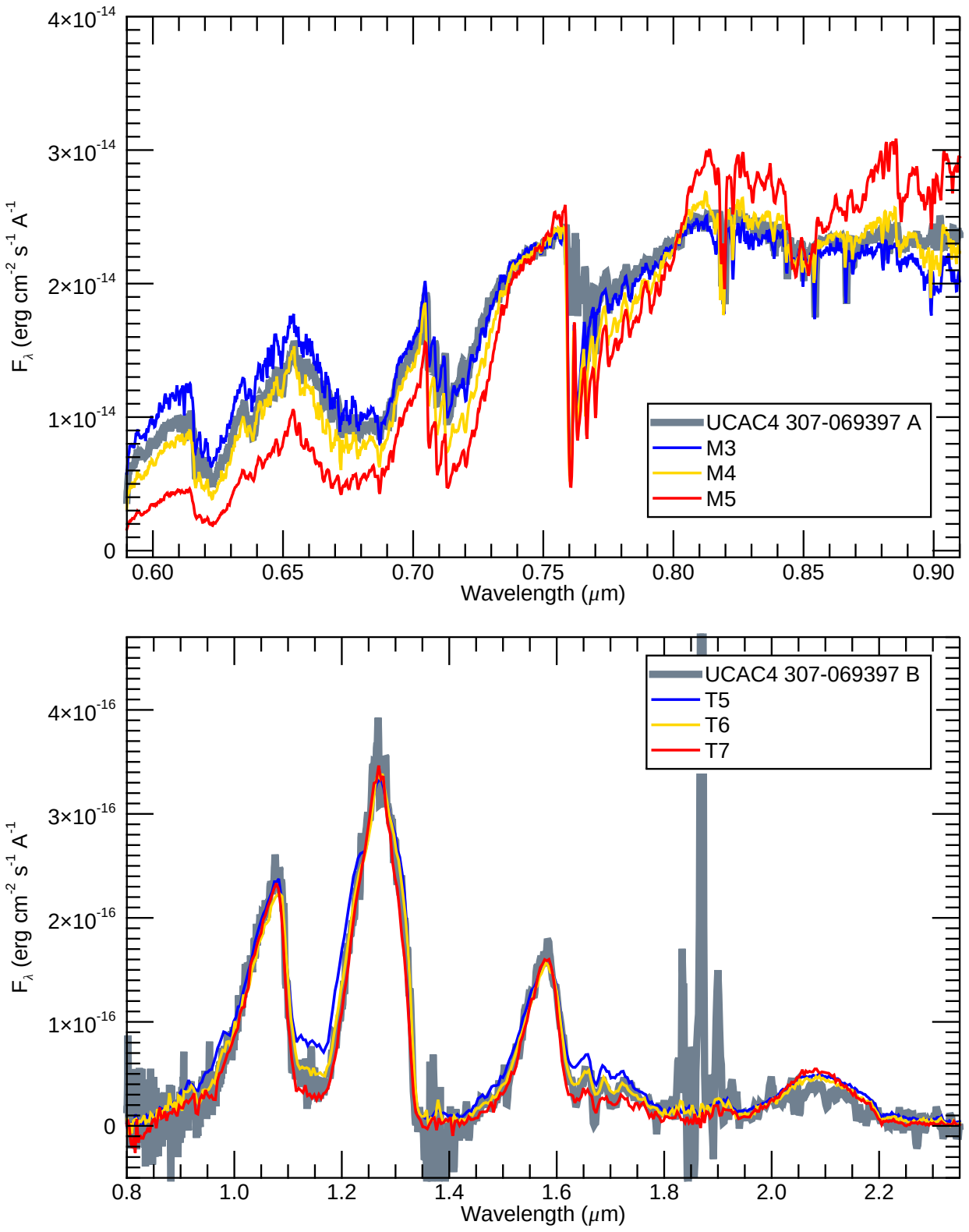

**Figure 14.** *Top:* the Lick/Kast spectrum of UCAC4 307-069397 A (1300A), plotted in slate grey. The M3, M4, and M5 spectroscopic standards are plotted in blue, gold and red. *Bottom:* The Magellan/FIRE spectrum of UCAC4 307-069397 B (1300B), plotted in slate grey. The T5, T6, and T7 spectroscopic standards are plotted in blue, gold and red. The target spectrum is smoothed with boxcar smoothing of width 3 pixels.

The companion is classified as T6 by comparison against template objects (Figure 14). The spectrum does not show clear peculiarities. Given the age constraints on the primary, the mass of the T6 is in the 47–64 $M_{\rm Jup}$ range.

### 8.11. *LP 270-10 AB (1353AB)*

The primary was observed by LAMOST (Zhao et al. 2012). We classify it as M2 V based on template matching, as illustrated in Figure 15. We note that we restrict the template comparison to the 0.6–0.8 μm range to avoid areas affected by known flux calibration issues (see e.g. Zhong et al. 2015). LAMOST DR5 reports an equivalent width for $H\alpha$ of 0.112 Å, consistent with our own measurement of 0.119±0.050 Å. Both measurements are well below the activity threshold established by Kiman et al. (2021). LAMOST DR5 claims a low metallicity of [Fe/H] = −1.19 dex, but visual inspection of the spectrum does not support this claim given the almost perfect match between the LAMOST spectrum and the solar-metallicity M2 V template. As an additional check, we measured the metallicity ourselves using the publicly available code am_getmetal[16] (Mann et al. 2013; Hawley et al. 2002). The code is designed to work on Supernova Integral Field Spectrograph (SNIFS, Lantz et al. 2004) or SpeX (Rayner et al. 2003) data, so we need to compensate for systematic differences due to different resolution and instrumental distortion. Therefore, we first searched for stars from Mann et al. (2013) in the LAMOST catalogue, finding 23 M dwarfs in common between the two. Then, we computed their metallicities with am_getmetal, and took the median difference between the values we compute using the LAMOST spectra and those published in Mann et al. (2013) as a measurement of the instrumental offset, and the one-sigma dispersion as an empirical estimate of our uncertainties. We found an offset of 0.21 dex and a dispersion of 0.24 dex. Our offset-corrected [Fe/H] measurement for 1353A is −0.11±0.24 dex, consistent with the solar-metallicity suggested by our template matching. The kinematic age constraint is 2.6–9.5 Gyr. The M3 V is detected by *GALEX* in the long-wavelength channel with NUV = 21.73±0.29 mag, but undetected in the short-wavelength channel implying FUV > 22.1 mag. The resulting $f_{\rm NUV}/f_{\rm J}$ places this object below the young M dwarf locus in Figure 3 of Shkolnik et al. (2011), and the NUV–W1 and J–W2 colors place it at the edge of the young M dwarf selection region defined by Rodriguez et al. (2013). Given the analysis above, we adopt an age in the range 2.6–9.5 Gyr for this object.

The companion likely has a spectral type of T6.5 based on its CatWISE2020 and *Spitzer* photometry and the polynomial relations of Kirkpatrick et al. (2021a). The CatWISE2020, *Spitzer*, and UHS photometry do not show any peculiarity, as expected given the disk age and solar metallicity of the system. Given the age constraints on the primary, interpolating the isochrones we obtain a mass in the range 40–60 $M_{\rm Jup}$.

### 8.12. *LP 81-30 AB (1416AB)*

We classify the primary as M2 V based on template matching of its LAMOST spectrum. The spectrum, along with the best fit templates, are shown in the top panel of Figure 16. The spectrum does not show any clear peculiarity. Using the LAMOST spectrum, we measure an $H\alpha$ equivalent width of 0.168±0.050Å, well below the activity threshold of Kiman et al. (2021). Using the same procedure described in Section 8.11, we measure [Fe/H] = –0.18±0.24 dex, in good agreement with the value reported by LAMOST DR5 (–0.19 dex). The *TESS* light curve is flat and featureless. 1416A is detected by *GALEX* in its NUV channel, with NUV = 20.99±0.26 mag, while it is undetected in the FUV channel, implying FUV>21.1 mag, consistent with older M dwarfs. The kinematic age range is 2.7–9.9 Gyr. The *Gaia* DR3 Astrophysical Parameters Inference System (APSIS, Creevey et al. 2022) reports an age of 3.1 Gyr, in good agreement with our kinematic age constraint. We adopt an age for 1416A in the 2.7–9.9 Gyr range.

[16] https://github.com/awmann/metal

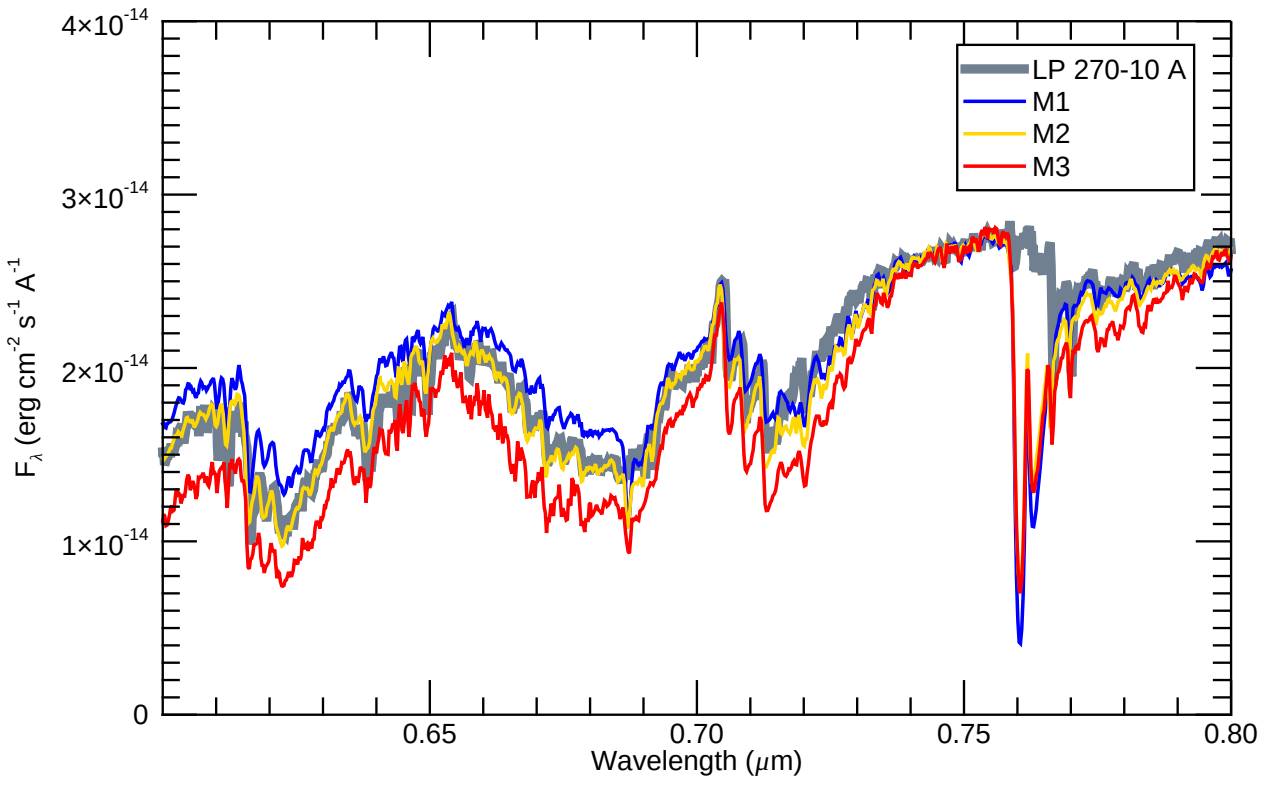

**Figure 15.** The 0.6–0.8 $\mu$m LAMOST spectrum of LP 270-10 A (1353A), plotted in slate grey. The M1, M2, and M3 spectroscopic standards are plotted in blue, gold and red, respectively.

The companion is classified T7 via template matching of its Keck/NIRES spectrum, shown in Figure 16. We note that the spectrum of the target shows some minor flux suppression in the K band with respect to the best-fit template. However, given that NIRES is a multi-order spectrograph, this could be due to an imperfection with our inter-order flux calibration and order merging. Given the measured spectral type and the age constraints from the primary, this T dwarf has a mass in the range 35–55 $M_{\rm Jup}$.

### 8.13. *G 135-35 AB (1417AB)*

The association between the two components of this system is questionable. Similarly to 0541AB, the two objects appear to be clearly comoving using WISEview. However, their measured proper motions are highly discrepant (see Table 4). The photometric distance to the T dwarf (∼43 pc, see below) and the astrometric distance to the M dwarf (∼28 pc) are also somewhat discrepant, leading to a comover probability of 0%.

However, *Gaia* reports a somewhat higher *ruwe* for the primary of 1.496, hinting at possible problems with the astrometric solution. Moreover, the T dwarf is the faintest in our sample, calling into question the reliability of its proper motion measurement. Dedicated astrometric observations for the companion are desirable to shed further light on the nature of this pair.

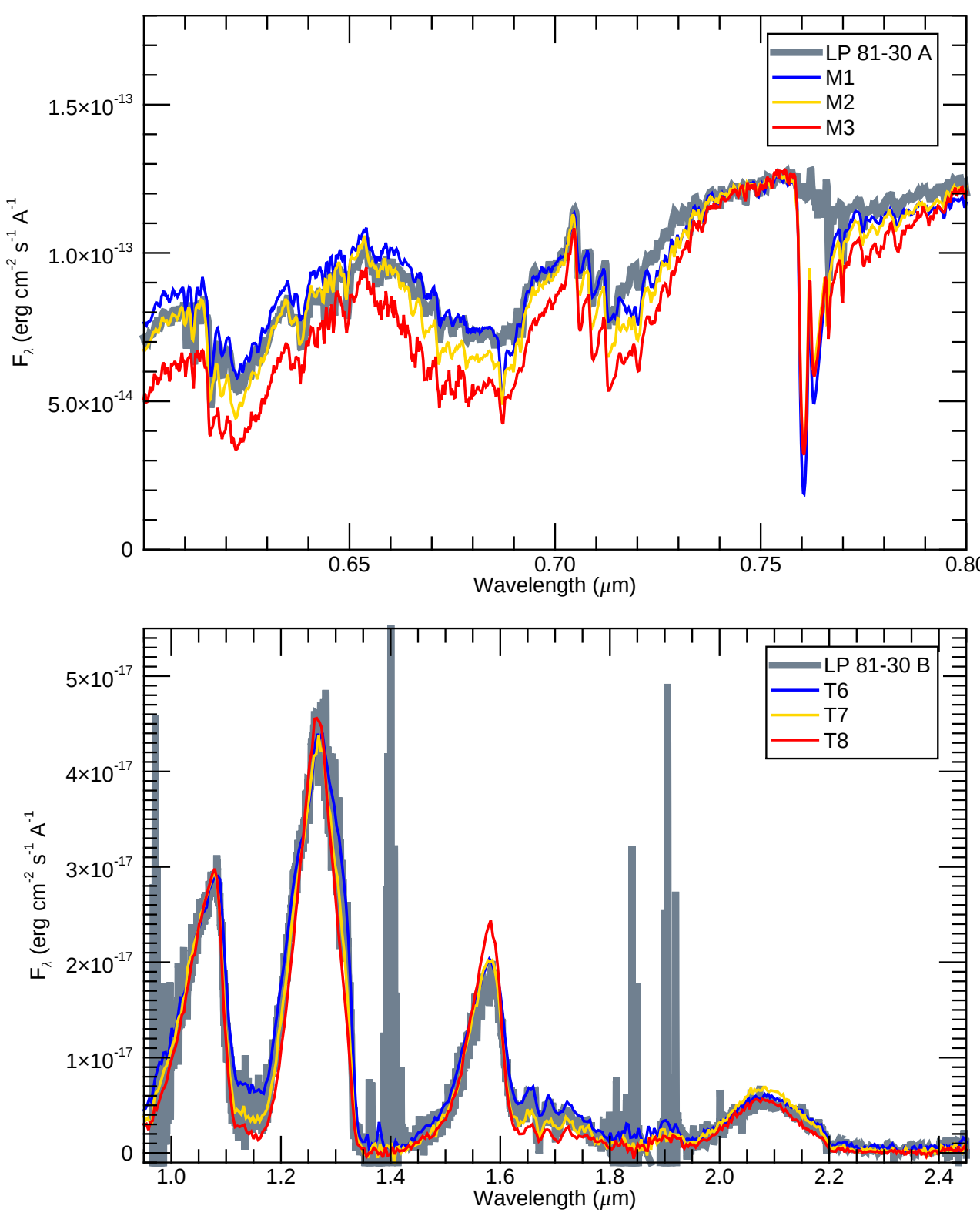

**Figure 16.** The LAMOST spectrum of LP 81-30 A (1416A; top) and the Keck/NIRES spectrum of LP 81-30 B (1416B; bottom), plotted in slate grey. In each panel, we plot spectroscopic standards in blue, gold and red. The spectrum of 1416B is smoothed with boxcar smoothing of width 5 pixels.

We classify the primary M3 V by fitting its SED (built using VOSA) with the M dwarf templates from Kesseli et al. (2017). The TESS 30-min light curve appears flat. The object is undetected in *GALEX*, implying NUV > 22.6 mag and FUV > 20.6 mag. The resulting NUV–W1 color limit and the $f_{\rm NUV}/f_{\rm J}$ limits place this M dwarf outside of the young stars loci defined in Rodriguez et al. (2013) and Shkolnik et al. (2011). The kinematic age is 4.1–11.7 Gyr and, as discussed in Section 7.4, the M dwarf has a probability of 99.5% of belonging to the thick disk. We assume the 4.1–11.7 Gyr age range as our best estimate for the age of the primary.

The companion is likely a T3 based on its CatWISE2020 photometry, implying a photometric distance of ∼43 pc. In the color-magnitude diagram of Figure 4, 1417B appears clearly underluminous and bluer than the bulk population of nearby T dwarfs. These characteristics are common among other known low-metallicity sdTs. This is consistent with the older age constraint and likely thick disk membership of the primary. If the association between the two objects is confirmed, the 1417AB system would be an important low-metallicity T dwarf benchmark. Spectroscopic follow-up is, therefore, desirable to further characterize 1417B. Given the age range for the primary, the mass of the companion would be in the 70–76 $M_{\rm Jup}$ range, just below the hydrogen burning limit.

## 9. SAMPLE PROPERTIES

In Section 8 we described how we used the age constraints derived from the M dwarf primaries to estimate the mass of the T dwarf companions.

We also estimated the mass of our M dwarf primaries using the $M_K$-to-mass relation of Mann et al. (2015), using the $K_s$ 2MASS magnitudes listed in Table 3 and the *Gaia* parallaxes. We then used these mass estimates, together with the mass constraints on the T dwarfs and the projected separation of the systems ($s$, see Table 1) to calculate the binding energy for each binary. It is important to remember at this point that the projected separation is not a direct measurement of the true semimajor axis $a$ of the orbit. Inclination angle, eccentricity, longitude of the ascending node, argument of periapsis, and true anomaly must be taken into account, so measuring $a$ requires full orbital reconstruction. Many papers use a correction factor of 1.26 to approximate $a$ given $s$ (i.e. $a = 1.26\ s$), however, this approximation assumes circular orbits. On the other hand, for wide binaries Tokovinin et al. (2020) performed numerical simulations assuming eccentric orbits with a thermal eccentricity distribution, and found that the median projected separation $s$ of the simulated population is an accurate measure of their median semimajor axis $a$, without the need for a correction factor. In any case, these considerations are only valid when comparing populations of objects, not when studying individual systems. Therefore, in our analysis we did not apply any correction factor.

With these caveats in mind, we compared our systems with stellar binaries and exoplanetary systems from the literature. Stellar binaries are taken from El-Badry et al. (2021), complemented with the list of low-mass systems compiled by Faherty et al. (2020) and Kiwy et al. (2022). Exoplanetary systems are taken from the NASA Exoplanet Archive. The top-left panel of Figure 17 shows the separation vs mass ratio parameter space, where we can see that nearly all of the new systems presented here are at the outskirts of the stellar binaries population. The young 0541AB system and the puzzling 0749AB system represent the more extreme outliers, occupying a scarcely populated region in this parameter space. The top-right panel of Figure 17 shows total system mass vs separation. The dashed line is the diffusive break-up limit determined by Weinberg et al. (1987) for system of 10 Gyr of age. According to Weinberg et al. (1987), systems to the right of this limit have an expected lifetime of less than 10 Gyr, as a result of galactic tides and catastrophic dynamical interactions with other stars and with giant molecular clouds. Once again, most of our systems fall in scarcely populated area at the limit of the stellar binaries population. The aforementioned 0541AB and 0749AB systems straddle the Weinberg stability limit, and it is perhaps not a coincidence that the only system beyond this limit, 0541AB, is young (however, see discussion on possible binarity of the primary in Section 8.6). These systems are particularly fragile and, therefore, unlikely to survive for very long in the Galactic disk. The bottom-left panel of Figure 17 shows binding energy vs. mass ratio. Several of the systems presented here clearly populate the gap between stellar and planetary systems, as they represent the most extreme stellar binaries. The bottom-right panel of Figure 17 paints a similar picture, with our newly discovered systems populating the extreme corner of the binding energy vs. total mass parameter space.

The growing population of wide, low-mass-ratio, low-total-mass binaries poses a challenge to the formation and evolution models of binaries and multiple systems and, therefore, represents a particularly important benchmark population to study and characterize.

## 10. SUMMARY

We have presented 13 new T dwarf companions in wide orbits around M dwarfs, identified using WISE data by the CatWISE and Backyard Worlds: Planet 9 projects. An object of particular interest is WISE 0751–7634, a previously known T9 (Kirkpatrick et al. 2011) which forms a common-proper-motion pair with L 34-26, a well studied young M3 V star within 10 pc of the Sun (Torres et al. 2006). The puzzling nature of this system poses a challenge to our current understanding of [Fe/H] and log $g$ effects on the spectra of T dwarfs. Long-wavelength spectroscopy with *JWST* is highly desirable to further understand the nature of this T9 companion. We also highlight CWISE J054129.32–745021.5 B, a T8 possibly associated with the very fast-rotating M4 V star CWISE J054129.32–745021.5 A. While the short rotation period of the primary hints at a very young age for this system, spectroscopic observations and the kinematics do not support this claim. Additional photometric and spectroscopic monitoring of CWISE J054129.32–745021.5 A is needed to better constrain the age of the system. Moreover, improved as-trometry for both components is needed to firmly establish the association of the two objects. Similarly, we have presented the 2MASS J05581644–4501559 AB and SCR J0959–3007 AB systems, where the M dwarf primaries show clear signs of youth. Their T dwarf companions are, therefore, directly-imaged planetary-mass objects, whose in-depth studies with *JWST* will provide important clues for a better understanding of giant exoplanets atmospheres. Finally, we mention UCAC3 52-1038 B, among the widest late T companions to main sequence stars, with projected separations of 7100 au. These very wide systems are challenging to model/explain for binary formation and evolution theory. Planet-like formation in-situ is impossible given the wide separation compared to the radii of typical proto-planetary disks, but formation close-in and subsequent migration also appears unlikely (Veras et al. 2009). Star-like formation would seem, therefore, the more probable explanation, but the survival of such a wide-separation low-mass companion in the dense environment typical of star-forming regions is also improbable (Weinberg et al. 1987).

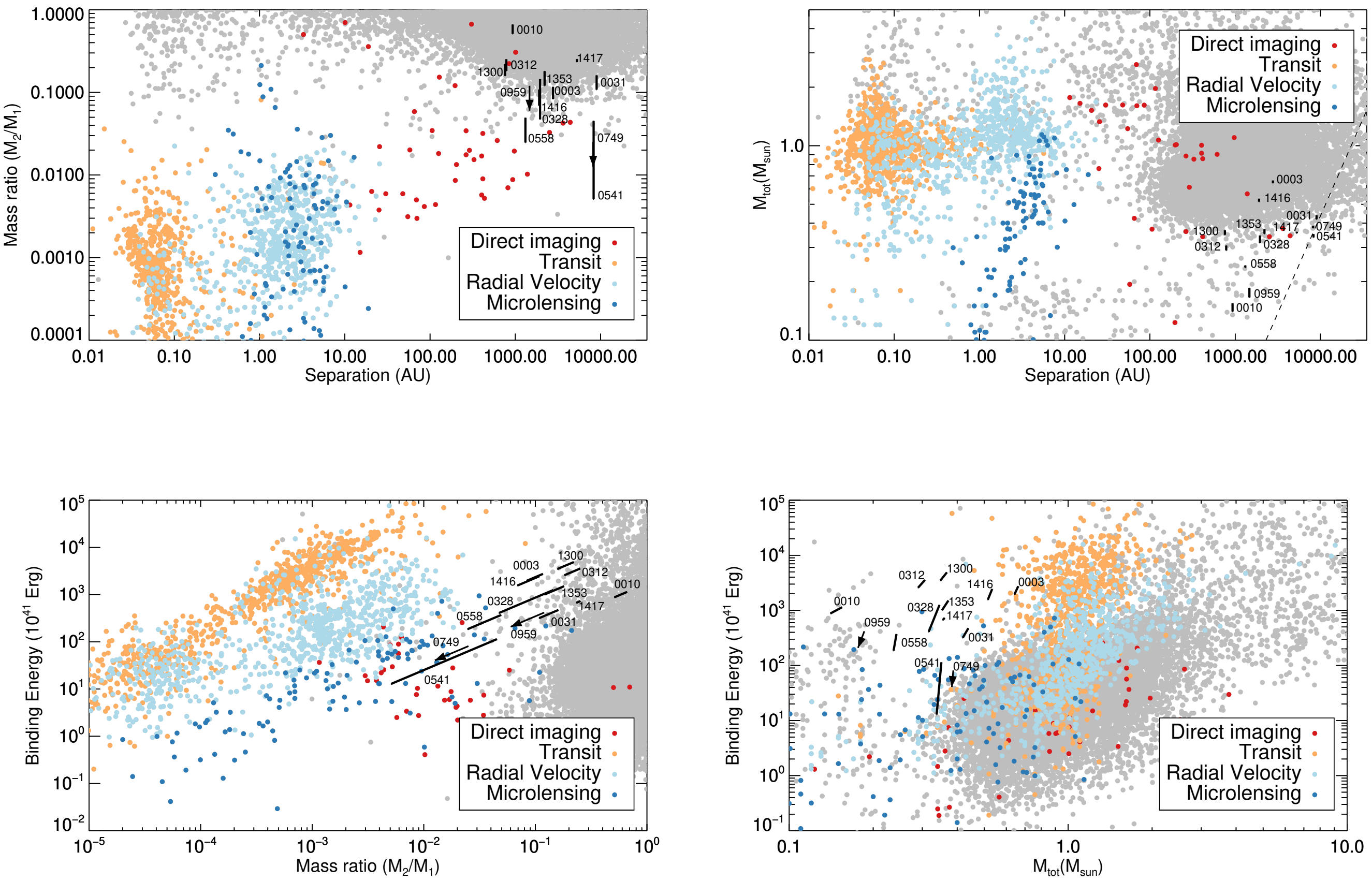

**Figure 17.** The new systems presented here, compared to the population of known stellar binaries (grey points) form Gaia eDR3 (El-Badry et al. 2021) complemented by the low-mass systems from Kiwy et al. (2022) and Faherty et al. (2020, and references therein). Also included for comparison is the exoplanets population from the NASA Exoplanet Archive (color-coded based on their discovery method). The range of properties for our targets are indicated by the individual black segments. The exceptions are the 0749AB and SCR J0959–3007 systems, for which we only have an upper limit on the mass of the companion (hence on the mass ratio and binding energy), which are plotted as arrows starting at the upper limit and pointing towards the direction corresponding to smaller secondary masses. The dashed line in the top-right panel is the diffusive breakup limit for a 10-Gyr system as defined by Weinberg et al. (1987). Our new binaries mostly occupy the mass ratio "gap" between stellar binaries and planetary systems, and are among the widest, lowest-total-mass systems known. As such, they tend to be "fragile" systems, and most of them are close to the breakup limit of the top-right panel.

The advent of the CatWISE2020 and *Gaia* catalogues, combined with the relentless work of enthusiastic citizen scientists, is revealing more of these wide star + brown dwarf binaries (Faherty et al. 2020; Schneider et al. 2021; Faherty et al. 2021; Rothermich et al. 2021; Kiwy et al. 2021; Gramaize et al. 2022; Kiwy et al. 2022). *EUCLID* (Euclid Collaboration et al. 2022) and future facilities (*NEOSurveyor*, Mainzer et al. 2023; *SPHEREx*, Alibay et al. 2023; *Rubin*, Ivezić et al. 2019) as well as the unprecedented follow-up capabilities offered by *JWST*, will enable even more discoveries and, therefore, more comprehensive, population-wide studies of this intriguing category of systems.

This work is based in part on observations made with the Spitzer Space Telescope, which is operated by the Jet Propulsion Laboratory, California Institute of Technology under a contract with NASA.

CatWISE was funded by NASA under Proposal No. 16-ADAP16-0077 issued through the Astrophysics Data Analysis Program, and uses data from the NASA-funded WISE and NEOWISE projects.

The Backyard Worlds: Planet 9 team would like to thank the many Zooniverse volunteers who have participated in this project. We would also like to thank the Zooniverse web development team for their work creating and maintaining the Zooniverse platform and the Project Builder tools. This research was supported by NASA grant 2017-ADAP17-0067. This material is supported by the National Science Foundation under Grant No. 2007068, 2009136, and 2009177.

This research made use of Lightkurve, a Python package for Kepler and *TESS* data analysis (Lightkurve Collaboration, 2018).

This publication makes use of VOSA, developed under the Spanish Virtual Observatory project supported by the Spanish MINECO through grant AyA2017-84089. VOSA has been partially updated by using funding from the European Union's Horizon 2020 Research and Innovation Programme, under Grant Agreement nº 776403 (EXOPLANETS-A).

NIRES data presented herein were obtained at the W. M. Keck Observatory, which is operated as a scientific partnership among the California Institute of Technology, the University of California, and the National Aeronautics and Space Administration. The Observatory was made possible by the generous financial support of the W. M. Keck Foundation. The authors recognize and acknowledge the significant cultural role and reverence that the summit of Maunakea has with the indigenous Hawaiian community, and that the W. M. Keck Observatory stands on Crown and Government Lands that the State of Hawai'i is obligated to protect and preserve for future generations of indigenous Hawaiians.

*Facilities:* WISE, Spitzer (IRAC), Keck (NIRES), Lick (Kast), SALT (RSS), Magellan (FIRE), Gemini (Flamingos-2), GALEX

*Software:* Spextool (Cushing et al. 2004); WISEView (Caselden et al. 2018); *lightkurve* (Lightkurve Collaboration et al. 2018); *wotan* (Hippke et al. 2019)

# APPENDIX

## A. FINDER CHARTS

In this appendix we present finder charts for the thirteen systems discussed in this paper. All finders are false-color *WISE* images, where W1 is in blue and W2 is in orange, and are centered on the *Gaia* coordinates of the primary. The finders were generated using WISEView (Caselden et al. 2018), coadding all NEOWISE images available for each field, which is typically 16 images spanning a time interval of 8 years (2014–2021).

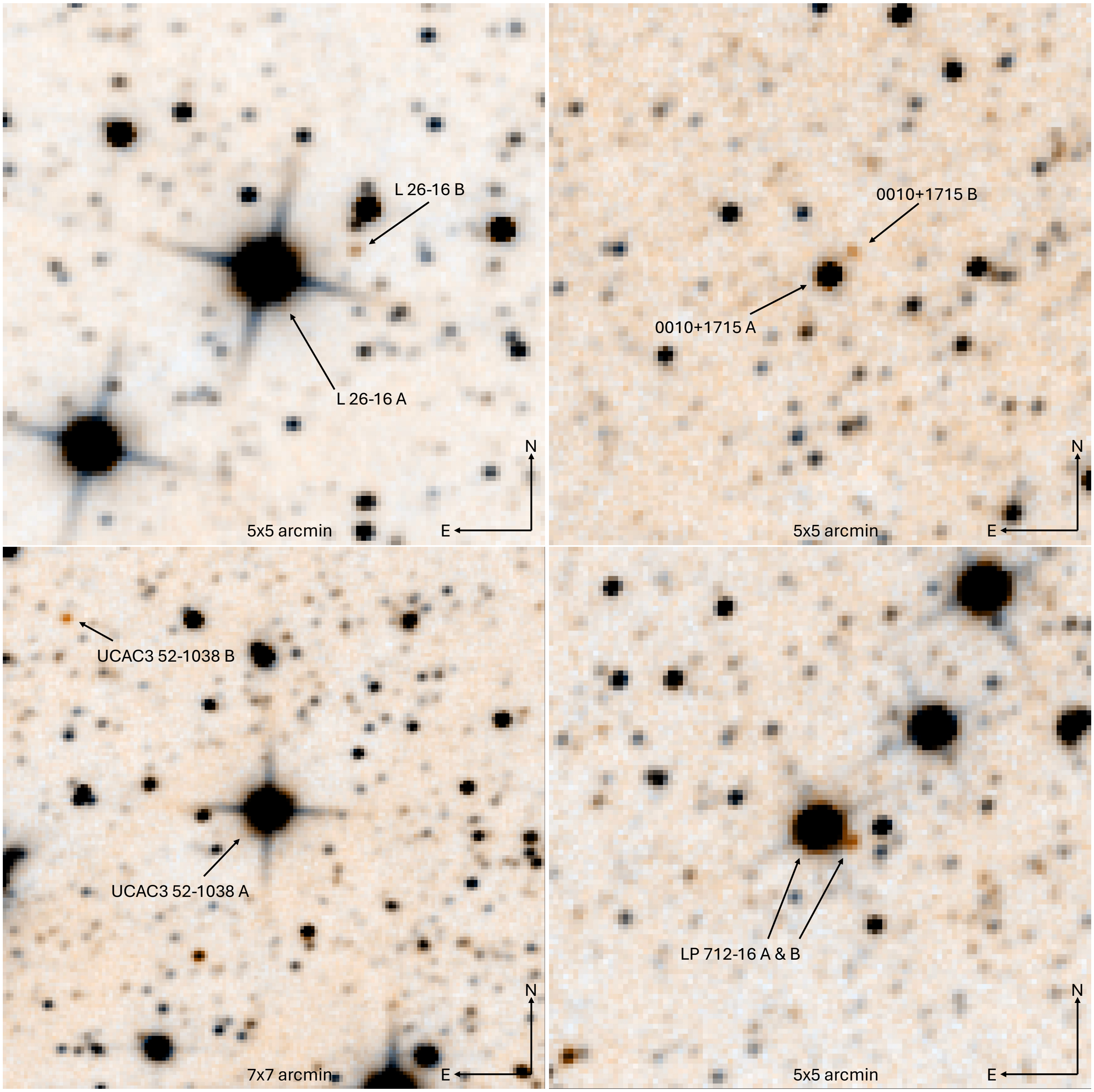

**Figure 18.** Finder charts for the L 26-16 AB system (0003AB; top-left panel), the 2MASS J00103250+1715490 AB system (0010AB; top-right panel), the UCAC3 52-1038 AB system (0031AB; bottom-left panel), and the LP 712-16 AB system (0312AB; bottom-right panel).

## B. TIME-RESOLVED T DWARF POSITIONS

The positions used for the proper motion measurements of the new T dwarfs presented here are given in Table 6. We refer the reader to Section 4 for further details on how the coordinates were measured, registered to the *Gaia* reference frame, and on how the proper motion for each object was measured.

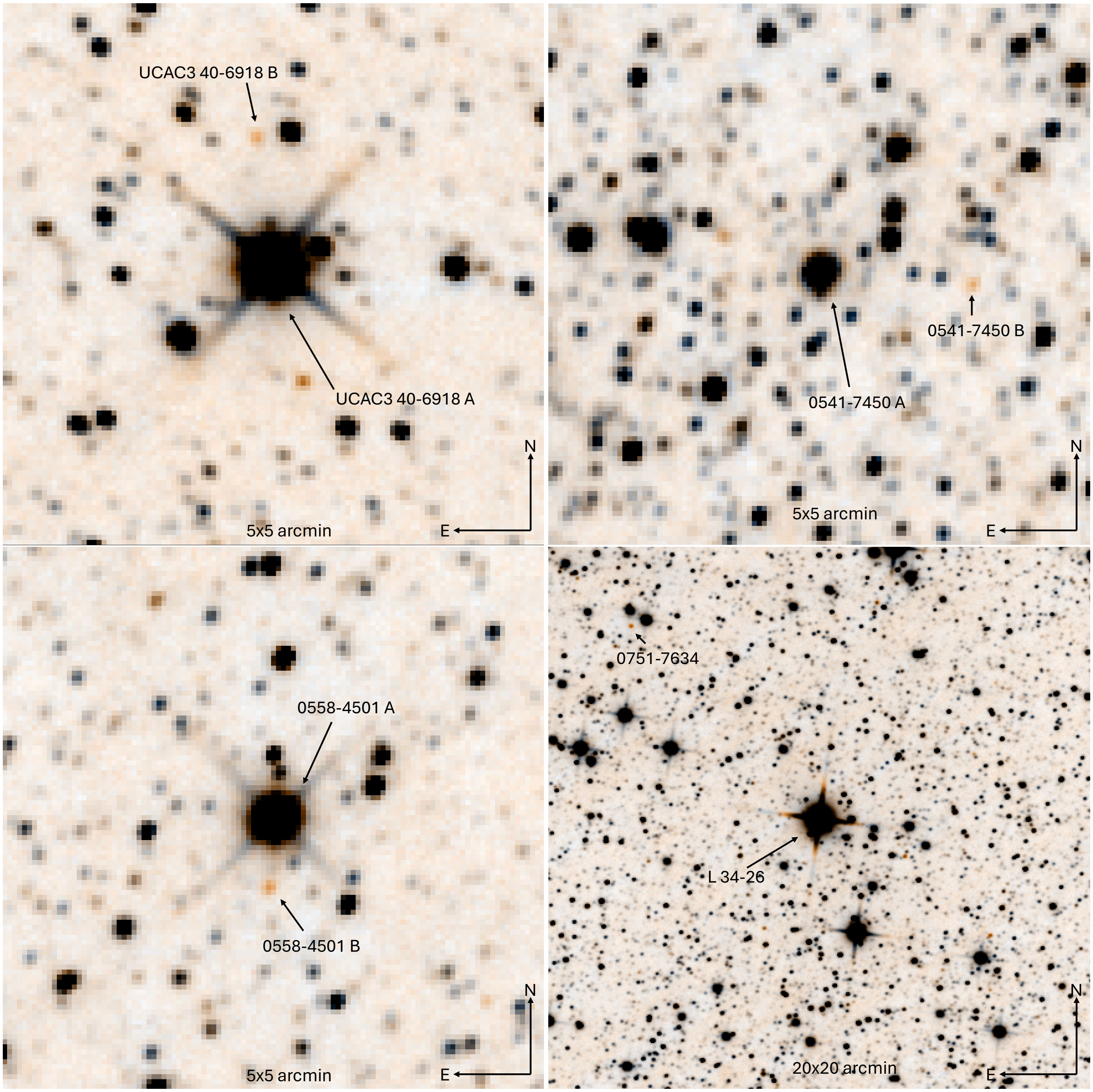

**Figure 19.** Finder charts for the UCAC3 40-6918 AB system (0328 AB; top-left panel), the CWISE J054129.32–745021.5 AB system (0541AB; top-right panel), the 2MASS J05581644–4501559 AB system (0558AB; bottom-left panel), and the L 34-26 + WISEP J0755108.79–763449.6 system (0749AB; bottom-right panel).

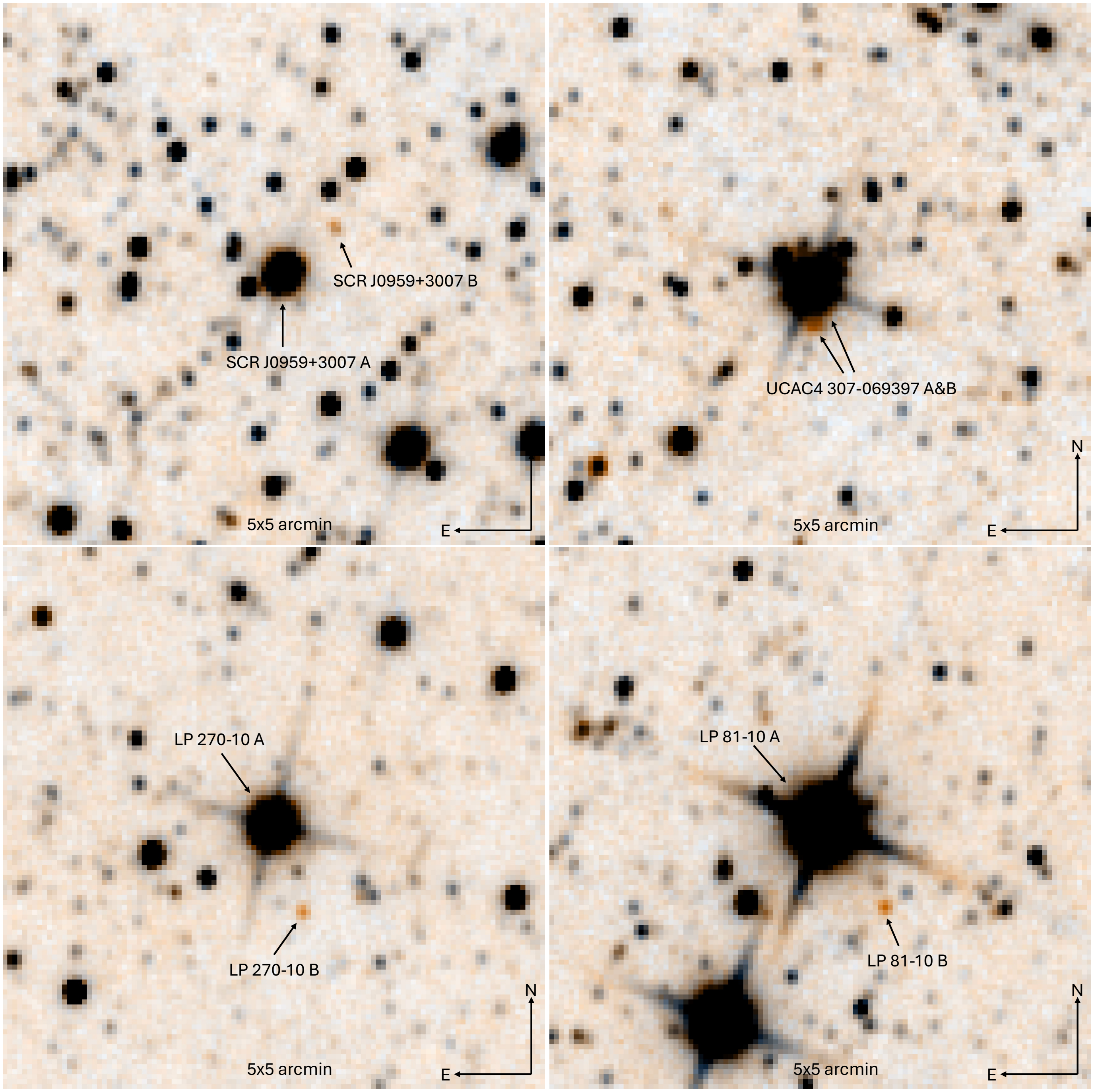

**Figure 20.** Finder charts for the SCR J0959-3007 AB system (0959AB; top-left panel), the UCAC4 307-069397 AB system (1300AB; top-right panel), the LP 270-10 AB system (1353AB; bottom-left panel), and the LP 81-30 AB system (1416AB; bottom-right panel).

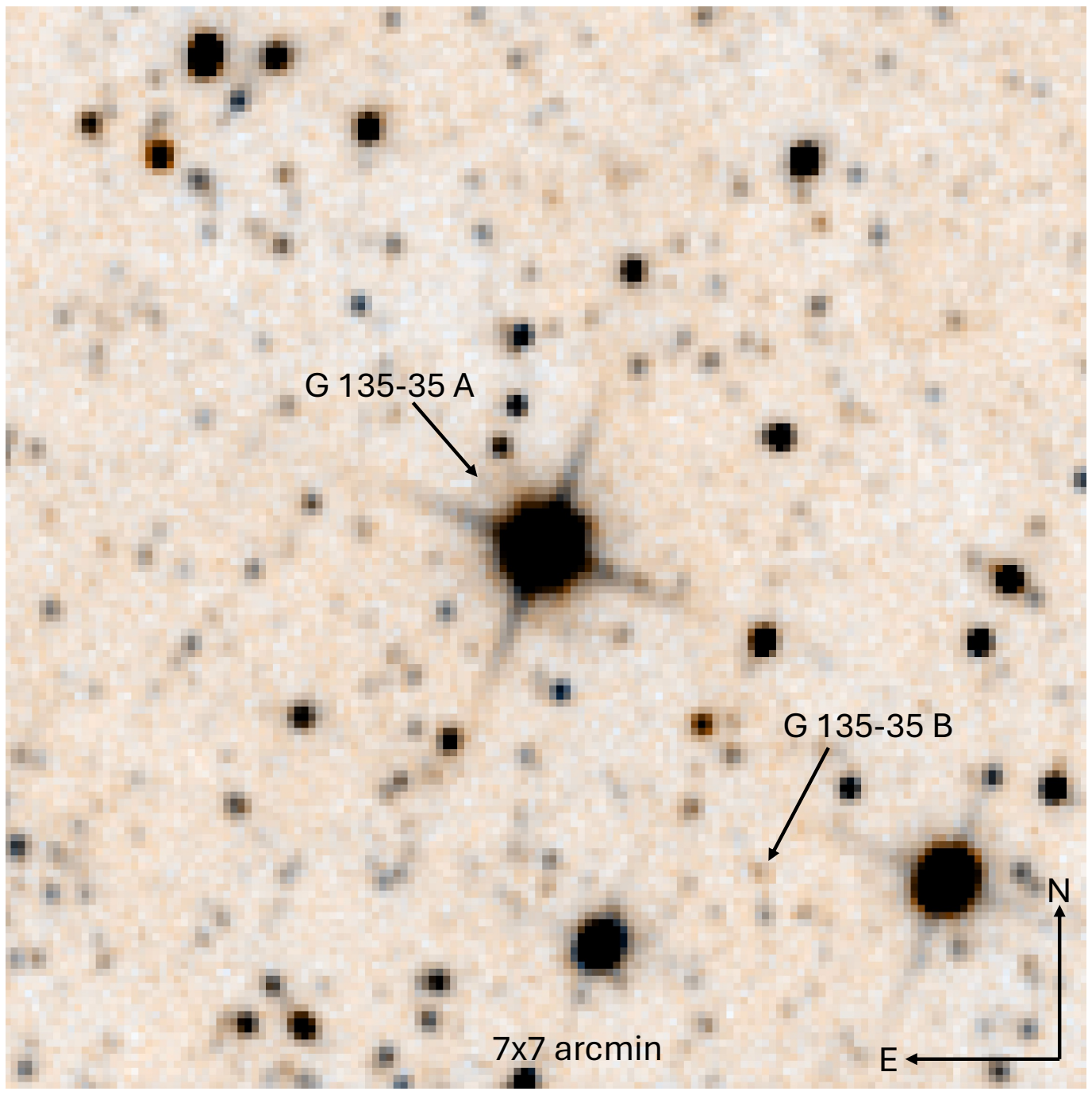

**Figure 21.** Finder chart for the G 135-35 AB (1417AB) system.

**Table 6**. Coordinates (J2000) used for the proper motion measurements presented in Table 4.

| ID | R.A. (deg) | $\sigma_{\mathrm{R.A.}}$ (arcsec) | decl. (deg) | $\sigma_{\mathrm{decl.}}$ (arcsec) | Epoch (yr) | Ref. |
|---|---|---|---|---|---|---|
| L26-16 B | 0.775537 | 0.44 | −75.552755 | 0.42 | 2010.308 | UW |
| L26-16 B | 0.775725 | 0.48 | −75.552774 | 0.44 | 2010.807 | UW |
| L26-16 B | 0.777529 | 0.46 | −75.552620 | 0.44 | 2014.316 | UW |
| L26-16 B | 0.776957 | 0.54 | −75.552651 | 0.51 | 2014.809 | UW |
| L26-16 B | 0.777956 | 0.55 | −75.552743 | 0.52 | 2015.309 | UW |

NOTE—Table 6 is published in its entirety in machine-readable format. A portion is shown here for guidance regarding its form and content. [a] *Spitzer* observations for this object were obtained as part of program 10046 (PI: Sanders). References: UHS = The UKIRT Hemisphere Survey DR1; VHS = The VISTA Hemisphere Survey DR5; SP = dedicated *Spitzer* observations (see Section 4); UW = unWISE time-resolved coadds (see Section 4).